\documentclass[12pt]{article}

\usepackage{theorem,amssymb,amsbsy,latexsym}

\newcommand{\vx}{{\bf x}}
\newcommand{\vy}{{\bf y}}
\newcommand{\prt}[1]{\bc \section*{{#1}} \ec}
\newcommand{\Lam}{\Lambda}
\newcommand{\remark}{\vspace*{0.45cm}

\noindent {\bf Remark: }}

\begin{document}
\begin{center}
{\Large \bf  Loop groups and quantum fields}\footnote{Published in: Geometric Analysis and Applications to Quantum Field Theory (Progress in Mathematics Vol 205), P. Bouwknegt and S. Wu (Eds.), Birkhauser, Boston (2002), 45-94}\\

\vspace{1 cm}

\vspace{.2 cm}
{\large Alan L. Carey}\footnote{Current email: {\tt acarey@maths.anu.edu.au}} \\
\vspace{0.2 cm}
{\em Department of Pure Mathematics, University of Adelaide,
Adelaide, South Australia 5005, Australia.\\
{\tt acarey@maths.adelaide.edu.au}  }\\
\vspace{.3 cm}
{\large and}\\
\vspace{.3 cm}
{\large Edwin Langmann}\footnote{Current email: {\tt langmann@kth.se}}\\
\vspace{0.2 cm}
{\em Theoretical Physics, Royal Institute of Technology, S-10044
Sweden.\\ {\tt langmann@theophys.kth.se} }\\
\vspace{.3 cm}

\vspace{0.2 cm}
%{\today}
Autumn 2001
\end{center}

\setcounter{footnote}{0}
\renewcommand{\thefootnote}{\arabic{footnote}}

\def\frac#1#2{{#1\over #2}}
\def\K{{\cal K}}
\def\M{{\cal M}}
\def\O{{\cal O}}
\def\cR{{\cal R}}
\def\CC{{\cal C}}
\def\CV{{\cal V}}
\def\W{{\cal W}}
\def\cZ{{\cal Z}}
\def\b{{\hat\rho}}
\def\ip#1#2{\langle #1,#2\rangle}
\def\tr{\rm tr}
\def\GR{Gr_{\rm tame}}
\def \G{\Map(\partial\Sigma, \U(1))}

%%%%%%%%%%%%%%%%%%%%%%%%%%%%%%%%%%%%%%%%%%%
\newcommand{\Map}{{\rm Map}}
\newcommand{\SU}{{\rm SU}}
\newcommand{\U}{{\rm U}}
\newcommand{\xx}{\stackrel {\scriptscriptstyle \times}{\scriptscriptstyle \times}}
\newcommand{\xxa}{\stackrel {\scriptscriptstyle \times}{\scriptscriptstyle \times} \!}
\newcommand{\xxe}{\! \stackrel {\scriptscriptstyle \times}{\scriptscriptstyle \times}}
\newcommand{\li}{\lim_{\eps\downarrow 0}}
\newcommand{\nt}{\int_{-L/2}^{L/2} dy\,}
\newcommand{\xxez}{\left.\xxe\right|_{\eps\downarrow 0}}
\newcommand{\f}{\frac}
\newcommand{\sgn}{{\rm sgn}}
\newcommand{\PiL}{\mbox{$\frac{\pi}{L}$}}
\newcommand{\zPiL}{\mbox{$\frac{2\pi}{L}$}}
\newcommand{\halfL}{\mbox{$\frac{L}{2}$}}
\newcommand{\half}{\mbox{$\frac{1}{2}$}}

\newcommand{\eq}{\begin{equation}}
\newcommand{\eqend}{\end{equation}}
\newcommand{\eqa}{\begin{eqnarray}}
\newcommand{\nonueqa}{\begin{eqnarray*}}
\newcommand{\eqaend}{\end{eqnarray}}
\newcommand{\nonueqaend}{\end{eqnarray*}}
\newcommand{\nonu}{\nonumber \\ \nopagebreak}
\newcommand{\bma}[1]{\begin{array}{#1}}
\newcommand{\ema}{\end{array}}
\newcommand{\bc}{\begin{center}}
\newcommand{\ec}{\end{center}}
\newcommand{\Ref}[1]{(\ref{#1})}

\renewcommand{\phi}{\varphi}
\newcommand{\eps}{\varepsilon}
\newcommand{\teps}{\tilde{\eps} }

\newcommand{\R}{{\mathbb R}}
\newcommand{\C}{{\mathbb C}}
\newcommand{\Z}{{\mathbb Z}}
\newcommand{\N}{{\mathbb N}}

\newcommand{\cA}{{\cal A}}
\newcommand{\cB}{{\cal B}}
\newcommand{\cC}{{\cal C}}
\newcommand{\cD}{{\cal D}}
\newcommand{\cL}{{\cal L}}
\newcommand{\cO}{{\cal O}}
\newcommand{\cP}{{\cal P}}
\newcommand{\cH}{{\cal H}}
\newcommand{\cF}{{\cal F}}
\newcommand{\cE}{{\cal E}}
\newcommand{\cG}{{\cal G}}
\newcommand{\cU}{{\cal U}}
\newcommand{\cS}{{\cal S}}
\newcommand{\cK}{{\cal K}}
\newcommand{\cT}{{\cal T}}
\newcommand{\cX}{{\cal X}}
\newcommand{\cW}{{\cal W}}
\newcommand{\cJ}{{\cal J}}

\begin{abstract}
This article surveys the application of the representation
theory of loop groups to simple models in quantum field theory
and to certain integrable systems.
The common thread in the discussion is the construction of
quantum fields  using vertex operators.
 These examples include the construction and solution
of the Luttinger model and other 1+1 dimensional interacting quantum field
theories, the construction of anyon field operators on the circle, the
`2nd quantization' of the Calogero-Sutherland model using anyons
and the geometric construction of quantum fields on Riemann
surfaces. We describe some new results on the elliptic Calogero-Sutherland
model.
\end{abstract}

\section{Introduction}
\label{s1}
The examples we discuss in this review
support the viewpoint ({\it cf} \cite{PS})
that much of $1+1$ dimensional quantum field theory
is the representation theory of
infinite dimensional groups. This is well understood
at the Lie algebra level
for conformal quantum field theory where the representation
theory of the Virasoro algebra and Kac-Moody algebras is central.
We take a somewhat different point of view in this
article and one that is more closely related to the
ideas generated by the representation theory of {\em loop groups} (see e.g.\
\cite{CR,F,Kac,KR,PS}) and the analysis of superselection
sectors \cite{BMT}. This view is  closely related to
 string theory and the analysis of
 integrable systems.  We spend some time surveying a number of
older developments and describe in detail
two newer applications. The first of these is the Calogero-Sutherland model
and its construction using anyon fields, while the second
reviews the properties of Fermion fields in Riemann surfaces
which is related to the study of the Landau-Lifshitz equation.
All of our examples have in common that they can be made
mathematically precise using an approach which is particularly simple
(even though we will only touch on those technical issues which are not
essential to the results).

Loop groups $\Map(S^1,G)$ are infinite dimensional Lie groups of smooth
maps from
the circle to some finite dimensional Lie group $G$ such as $\SU(n)$ or
$\U(n)$. We will mostly restrict ourselves to the
simplest case $\Map(S^1,\U(1))$.  We will review some of the important
results from the projective representation
 theory of this group and
show how these results are applied to
models in quantum field theory.  Especially interesting
are applications which yield new information, unattainable in a
precise mathematical form in any other way. The most interesting
use of these methods is in constructing
the solutions of integrable systems.
To make the later discussion accessible we begin in Part~A with
the construction and solution of the {\em
Luttinger model}, a simple model for interacting fermions in 1+1 dimensions
which, in a rather delicate limit, reduces to the massless Thirring model.
We will then discuss modifications of the method which leads to
the Luttinger model and show how other examples can be obtained
with a particular emphasis on fields with braid group statistics.

Brief descriptions of these other models are contained
in Subsection~3.2 and in Part~B, Subsection~6.1 with references
to the literature.
{}From the viewpoint of conformal field theory
these are mostly genus zero examples.
In Part~B we discuss
the geometric viewpoint on free fermions
at non-zero temperature and find that they
provide a genus 1 example i.e. fields
on a torus. This leads into a discussion of fermion
field theories on higher genus Riemann surfaces.
We find that associated to these are representations
of `generalised' loop groups.
(These apply, in the spirit of
the other applications discussed in this review,
to the construction of the soliton solutions of the Landau-Lifshitz
equation which we briefly describe).

The latter portion of Part~A is a review of our recent paper \cite{CL},
with some extensions, on
constructing quantum fields that are neither bosons nor fermions
but satisfy
non-trivial algebraic `exchange relations'.
Crucial to this is the existence of a {\em
boson-anyon correspondence} which shows that the fields satisfying
braid statistics (the so-called
anyon field operators) can be
obtained as a limit of
operators representing certain special elements of
the loop group. This result
allows explicit computation of all anyon correlation functions.

To be specific, {\em anyons on the circle} are quantum field
operators obeying exchange relations
\eq
\label{any}
\phi^\nu(x)\phi^{\nu'}(y) = e^{i\pi\nu\nu'\sgn(x-y) }\phi^{\nu'}(y)\phi^\nu(x)
\quad \forall x,y\in S^1, x\neq y .
\eqend
For $\nu^2$ and $(\nu')^2$ both even (odd) integers these fields are bosonic
(fermionic), but we are interested in {\em any}ons where $\nu$ and $\nu'$
can be {\em any} real numbers.  The basic idea of how to generalize the
boson-fermion correspondence to anyons is
quite old, see e.g.\ \cite{Klaiber},
provided one is not too concerned with the
exact mathematical meaning of the construction.
However, there are delicate technical
points related to the distributional nature of quantum fields.
We will show how the theory of loop groups allows us to handle these
difficulties quite elegantly.  Our method can construct anyon field
operators satisfying Eq.\ \Ref{any} only if $\nu$ and $\nu'$ are integer
multiples of some fixed $\nu_0\in\R$.
These anyon field operators can be used to solve the {\em
Calogero-Sutherland model} \cite{Su} which is defined by the
Hamiltonian
\eq
\label{Sutherland}
H_{N,\nu^2} = -\sum_{k=1}^N \frac{\partial^2}{\partial x_k^2}
+\sum_{1\leq k<\ell\leq N } 2 \nu^2(\nu^2-1) V(x_k-x_\ell)
\eqend
with $-L/2\leq x_j\leq L/2$ coordinates on a circle of length $L$,
$\nu>0$, $N=2,3,\ldots$, and
\eq
V(r) = -\f{\partial^2} {\partial r^2}  \log\sin (\PiL r)
\eqend
i.e.\ $V(r) = (\PiL)^2 \sin^{-2}(\PiL r)$.
This is an integrable quantum mechanical model of $N$
interacting particles moving on a circle.  This model has
received considerable attention recently, especially in the context of the
quantum Hall effect and conformal field theory, see e.g.\
\cite{AMOS1,AMOS2,Iso,MS}.

The idea of \cite{CL} is to
generalize the construction of the wedge representation of the
$W_{1+\infty}$-algebra (see e.g.\ \cite{KR0}), which uses fermions, to
anyons.  As a motivation of this, we discuss in some detail how,
in the fermion case,
a `generating function' for the operators representing the Abelian subalgebra
of the $W_{1+\infty}$-algebra can be obtained as a simple application of
loop group theory.  These operators $W^{s+1}$, $s=0,1,2\ldots$ generalize
the fermion charge ($s=0$) and the free fermion Hamiltonian ($s=1$) in a
natural manner.  They are all local and quadratic in the fermion fields,
and we obtain alternative representations which are local and in powers
of boson fields.  This corresponds to a generalization of
the Sugawara constructions (see e.g.\
\cite{GO}), and we refer to these as {\em generalized Kronig identities}
(which is the special case $s=1$). Remarkably a
further generalisation to
anyons with an arbitrary statistics parameter $\nu$ is possible
and which is straightforward only for $s=0,1$.  The first
non-trivial case is $W^{3}$ which can be regarded as a second
quantization of the Calogero-Sutherland Hamiltonian with the coupling constant
determined by the statistics parameter $\nu$.
The anyon-analog of $W^{3}$ obeys exchange relations with products
of anyon field operators which may be exploited to construct
iteratively, eigenfunctions of the
Calogero-Sutherland model and thus recover the solution of this
model \cite{CL} found originally in \cite{Su}.

We end Part~A with an outline of a further generalization of this
construction in which anyons at a finite temperature $1/\beta$
are constructed and used to find a second quantization of the elliptic
generalization of the Calogero-Sutherland model in which the interaction
potential $V(r)$ is equal to the Weierstrass elliptic function $\wp(r)$
with periods $L/2$ and $i\beta L/(2\pi)$ \cite{L4}.

\subsection{Summary} This review starts in
Section~\ref{s2} with the projective representation theory of loop
groups based on the quasi-free representations of fermion field
algebras. This is the so-called wedge representation of the loop
group.  In this Section we also review the boson-fermion
correspondence.  In Section~\ref{s3} we outline how these results are
used to treat the Luttinger model.  Our discussion of the
$W_{1+\infty}$ algebra is contained in Section~\ref{s4}.
Section~\ref{s5} explains the results on the Calogero-Sutherland model and
anyons.  Part~B begins with an overview of other two dimensional
quantum field theories which can be constructed using representations
of loop groups (Section~\ref{s6}). In Section~\ref{s7} we revisit the
theory of free fermions at non-zero temperature and show how these
fields may be interpreted as living on a torus.  (The exposition
follows \cite{CH1}. The result is, however, folklore.) This motivates a
discussion of fermions on Riemann surfaces in Section~\ref{s8},
synthesising examples and ideas from \cite{Seg2,CEH,CH1,CHM,CHMS,CHa}
but mainly following \cite{CH2}.  The main point of our exposition is
to sketch how the geometry of a Riemann surface with boundary
determines a representation of an associated infinite dimensional
group of $\U(1)$-valued functions on the boundary.  We then focus on
the construction of vertex operators on Riemann surfaces by
generalising the construction of Part~A.

The discussion of the quantum field theory applications is based on
many papers \cite{CEH}--\cite{CW},\cite{GL1}--\cite{GLR},
\cite{L1,L4,LS} with emphasis on \cite{GLR,CL,CH1,CH2}.  Most of
Section~\ref{s4} and Subsection~5.2.2 are new.  A portion of
Part~A appeared in \cite{LC}
and a recent pedagogical introduction to some of the material
described here is given in \cite{L8}.

In the present review we restrict ourselves to quasi-free second
quantization of fermions and two dimensional quantum field theory
models.  We only mention in passing that bosons can be treated in a
similar manner, see e.g.\ \cite{Ruij,L1}, and that there is a
super-version of quasi-free second quantization in which bosons and
fermions are treated simultaneously and on the same footing using
$\Z_2$-graded algebric structures \cite{GL1,GL2}.  We also mention
that there is an interesting relation between quasi-free second
quantization and Connes' noncommutative geometry \cite{Connes} (a
recent textbook is \cite{GVF}) and higher dimensional quantum gauge
theories. Some of these developments were reviewed in \cite{L3} (see
also \cite{GVF}).

\prt{PART A: LOOP GROUPS, FERMIONS AND PHYSICS}

\section{Loop groups and quantized fermions}
\label{s2}
In this section we review some mathematical results on loop groups and
quasi-free representations of fermion field algebras which will play a
central role in the following.  The material is standard, see e.g.\
\cite{Ar,Kac,KR,M2,PS}. We follow mainly the discussions in \cite{CR} and
\cite{Seg1}.

\subsection{Notation}
\label{s2.1}
Throughout this part, $x\in [-L/2,L/2]$ is a coordinate on a circle of
length $L$ which we denote as $S_L^1$.  Let $\cG=\Map(S_L^1;\U(1))$ be the
set of smooth maps $S_L^1\to\U(1)$.  We note that each loop $\phi\in \cG$ can
be written as
\eq
\label{f}
\phi(x)=e^{i f(x)},\quad f(x)=w \f{2\pi}{L}x + \alpha(x)
\eqend
where $w = [f(L/2)-f(-L/2)]/2\pi$ is an integer called the {\em winding
number}, and $\alpha$ is a smooth map $S_L^1\to\R$.  We will find
it convenient to decompose such maps into positive-, negative- and
zero Fourier components,
\eqa
\label{alpha}
\alpha(x)=\alpha^+(x)+\alpha^-(x)+\bar\alpha\nonu
\alpha^\pm(x) =
\frac{1}{L}\sum_{\pm p>0} \hat\alpha(p) e^{ipx}, \quad
\bar\alpha = \frac{1}{L} \hat\alpha(0)
\eqaend
where we use the following conventions for Fourier transformation of loops,
$$
\hat\alpha(p)= \int_{-L/2}^{L/2} dx \, \alpha(x) e^{-ipx}\quad
p\in\Lambda^*
$$
where
$$
\Lambda^* := \left.\left\{ p=\frac{2\pi}{L}n   \right| n\in \Z\right\}.
$$

\subsection{Loop group of maps $S_L^1\to\U(1)$}
\label{s2.2}
We note that $\cG$ is a Lie group under point-wise
multiplication, $(\phi_1\cdot \phi_2)(x)=\phi_1(x)\phi_2(x)$.
It is known that $\cG$ has an interesting central extension
$\widehat\cG=\U(1)\times \cG$ with the group
multiplication
\eq
\label{p}
(\gamma_1,\phi_1) \cdot (\gamma_2,\phi_2) :=
(\gamma_1\gamma_2 \sigma(\phi_1,\phi_2), \phi_1\cdot\phi_2 )
\eqend
[$\gamma_i\in\U(1),\; \phi_i\in\cG, $] where
$$
\sigma( e^{if_1}, e^{if_2} ) = e^{-  i S (f_1,f_2)/2} ,
$$

\eqa
\label{S} S(f_1,f_2) = \frac{1}{4\pi}\left[ f_1\left(\halfL\right)f_2\left(-\halfL\right)
- f_1\left(-\halfL\right)f_2\left(\halfL\right) \right] + \nonu +
\frac{1}{4\pi} \int_{-L/2}^{L/2} dx
\left(\frac{d f_1(x)}{dx}f_2(x) -f_1(x) \frac{d f_2(x)}{dx}\right)
\eqaend
is a {\em two cocycle} of the group $\cG$: it satisfies
$$
\sigma(\phi_1,\phi_2)\sigma(\phi_1\cdot\phi_2,\phi_3)=
\sigma(\phi_1,\phi_2\cdot\phi_3)\sigma(\phi_2,\phi_3)
$$
which is equivalent the associativity of the group
product defined in Eq.\ \Ref{p}.

Below we will describe in some detail the construction of the so-called
{\em wedge-represen\-tation} of $\widehat\cG$, $(\gamma,\phi)\to
\gamma\Gamma(\phi)$, on the fermion Fock space $\cF$ over $L^2(S_L^1)$. The
$\Gamma(\phi)$ are unitary operators satisfying
\eq
\label{sigma}
\Gamma(\phi_1)\Gamma(\phi_2)=\sigma(\phi_2,\phi_2)\Gamma(\phi_1\cdot\phi_2)
\eqend
and
\eq
\label{Gammastar}
\Gamma(\phi)^*=\Gamma(\phi^*)
\eqend
(for simplicity in notation, we denote the Hilbert space adjoint and
complex conjugation by the same symbol $*$).  Moreover, there is a vector
$\Omega\in\cF$ such that for all $f$ of the form Eq.\ \Ref{f},
\eq
\label{delta}
\left<\Omega,\Gamma(e^{if})\Omega\right> = \delta_{w,0} e^{-i S(\alpha^-,\alpha^+) }
\eqend
where $<\cdot,\cdot>$ is the inner product in $\cF$. Note that
$$
i S(\alpha^-,\alpha^+) = \sum_{p>0}
\frac{p}{2\pi L} \hat\alpha(-p)\hat\alpha(p)
$$
is positive definite.

We now describe how this representation $\Gamma$ of $\widehat\cG$ is
constructed.

\subsection{Quasi-free second quantization of fermions}
\label{s2.3}
\subsubsection{Fermion field algebras}
Let $\cH$ be a separable Hilbert space.  The {\em fermion field algebra} $\cA$
over $\cH$ is then defined as the $C^*$-algebra generated by elements
$a^*(f)$ and $a(f)=a^*(f)^*$ such that $f\to a^*(f)$ is linear,
$||a^*(f)||^2=\langle f, f\rangle_{\cH}$, and the canonical anticommutation
relations (CAR) hold,
\eq
\label{CAR}
a(f)a(g)+a(g)a(f)=0,\quad a(f)a(g)^*+a(g)^*a(f)=
\langle f, g\rangle_{\cH} I
\eqend
(here and in the following, $I$ denotes the identity operator).
The Fermion Fock space $\cF$ over $\cH$
is the Hilbert space obtained by completing
the exterior algebra $\wedge \cH$ over $\cH$ in the obvious Hilbert space
topology.  We define an action of $a(g)^*$ by
\[
   a(g)^* g_1 \wedge g_2\wedge\ldots\wedge g_n=g\wedge g_1\wedge
g_2\wedge\ldots\wedge g_n
\]
for $g_j$ in $\cH$.  Then $a(g)$ may be identified with the
Hilbert space adjoint of $a(g)^*$ and
it is easy to see that the anti-commutation relations \Ref{CAR} hold.
In this way one obtains the so-called {\em Fock-Cook representation}
of the fermion field algebra $\cA$.

\remark In applications to models in physics $\cH$ is taken as the
Hilbert space of 1-particle states. For example for
statistical mechanical models of fermions with
spin on a {\em finite} lattice $\Lam$, the 1-particle states are
$\C^2$-valued function on $\Lam$, i.e.\ $\cH \cong \C^{2|\Lam|}$ is
actually finite dimensional. In this case the  algebra $\cA$ and the
Fock space $\cF$ are also finite dimensional,\footnote{One can check that
$\dim_{\C}(\cF) = 2^{\dim_{\C}(\cH))}$ if $\dim_{\C}(\cH)<\infty$.}
all the irreducible representations are $\cA$ are unitarily equivalent,
and all one ever needs is the Fock-Cook representation described above.
The situation becomes mathematically more interesting if $\cH$ is
infinite dimensional. This is the situation for quantum field models
on a continuous manifold $M$ where the appropriate 1-particle space
$\cH$ is typically a space  of square integrable functions
on $M$.

\subsubsection{Quasi-free representations I. Irreducible case}
Let $P_-$ be a projection operator on $\cH$ (i.e.\
$P_-^2=P_-^*=P_-$) and let $P_+= 1-P_-$.
Then there is a representation $\pi_{P_-}$ of $\cA$ on
the fermion Fock space $\cF$ over $\cH$ which is determined by the
following conditions,
\eq
\label{qf}
\psi(P_+f)\Omega=0=\psi^*(P_-f)\Omega \quad \forall f\in\cH
\eqend
where we
write $\psi(f)=\pi_{P_-}(a(f))$; $\Omega$ is the cyclic (or vacuum) vector
in the $\cF$.
One can prove that the representations $\pi_{P_-}$ are
irreducible \cite{BR}.

\remark In applications, $P_-$ usually is determined by
a self-adjoint operator $D$ on $\cH$ which represents the
1-particle Hamiltonian (i.e.\ energy operator)
of a specific model and typically is some Dirac operator
with a spectrum which is unbounded from above and below.
In this situation $P_-$ is taken as the spectral projection
of $D$ corresponding to the interval $(-\infty, 0)$
i.e.\ $P_- = \theta(-D)$,
where $\theta(x)=1$ for $x \geq 0$ and $\theta(x)=0$
for $x<0$. Then the state $\Omega$ above
corresponds to the so-called {\em filled Dirac sea} and is the
ground state (i.e.\ state of least energy) of the many
particle Hamitonian corresponding to $D$, as explained in
more detail below.

\subsubsection{Second quantization of 1-particle operators}
Let $g_1$ the set of all bounded operators $X$ on
$\cH$ such that $P_\pm X P_\mp$ is Hilbert-Schmidt. $g_1$
is a Lie algebra, and for each $X\in g_1$ there is operator
$d\Gamma(X)$ acting on $\cF$ such that \cite{CR}
\eq
\label{cc}
[d\Gamma(X),\psi^*(f)] = \psi^*(Xf)\: ,
\eqend
\eq
\label{ccc}
d\Gamma(X)=d\Gamma(X^*)^* \: ,
\eqend
and
\eq
< \Omega,d\Gamma(X)\Omega >_{\cF}\, =0 \: .
\eqend
The construction of these operators requires a regularization ---
physicists refer to it as `normal ordering' --- and due to this, $X\to
d\Gamma(X)$ is not a representation but rather a {\em projective
representation} of $g_1$: One has relations
\eq
\label{cc1}
[d\Gamma(X),d\Gamma(Y)] = d\Gamma([X,Y]) + i\hat S(X,Y) I
\eqend
where
\eq
\label{cc12}
i\hat S(X,Y) = {\rm Trace}_{\cH}(P_-XP_+YP_- - P_-YP_+XP_-)
\eqend
is a {\em non-trivial two cocycle} of the Lie algebra $g_1$ \cite{CR,Lund}.
In the physics literature $i\hat S$ is known as the {\em Schwinger term}.
For $X\in g_1$ one also has
\eq
\label{cc2}
P_+ X P_- = 0 \; \Rightarrow d\Gamma(X)\Omega = 0
\eqend
which is called {\em highest weight condition}.

We note that even for bounded $X$,
the operators $d\Gamma(X)$ are unbounded, but one can easily
construct a common dense invariant domain on which all
the relations above are
well-defined \cite{CR,GL2}.  Moreover, the construction of $d\Gamma(X)$ can
be naturally extended to certain algebras of {\em unbounded} operators
$X$ on $\cH$ which have a common dense invariant domain of definition
\cite{GL2}.  All relations given above naturally extend to this larger Lie
algebra of operators on $\cH$.  In the following the same symbol $g_1$ will
be used also for such Lie algebras of unbounded operators on $\cH$.

Let $G_1$ be the set of all unitary operators $U$ on $\cH$ with $P_\pm
UP_\mp$ Hilbert-Schmidt.  This is a Lie group with a Lie algebra containing
the self-adjoint operators in $g_1$.  It has a projective representation
$U\to \Gamma(U)$ on $\cF$ such that
\eq
\Gamma(U)\psi(f)\Gamma(U)^{-1}= \psi(Uf).
\eqend
We say that $\Gamma(U)$ {\em implements} $U$.
`Projective' here means that relations
\eq
\Gamma(U)\Gamma(V) = \hat\sigma(U,V) \Gamma(UV)
\eqend
hold with $\hat \sigma$ a non-trivial phase factor.
An explicit formula for this cocycle $\hat \sigma$ was derived
in \cite{L1}.\\

\remark In a specific model, 1-particle observables are given
by self-adjoint $X$ on $\cH$ and $d\Gamma(X)$ (if it exists)
represents the corresponding many particle
observable. For example, if $D$ is the
1-particle Hamiltonian then $d\Gamma(D)$ is the many particle
Hamiltonian. One can show that in the quasi-free representation
$\pi_{P_-}$ with $P_- = \theta(-D)$, $d\Gamma(H)$ is always
positive. This is the reason why the quasi-free representations
are needed. One essential physical requirement in every
quantum model is the existence of a {\em ground state} i.e.\
state of lowest energy. If the 1-particle Hamiltonian is not bounded
from below then there is no groundstate, neither in the 1-particle
Hilbert space nore in the Fock-Cook representation, and therefore
they both have to be rejected. On the other hand, a representation
in which the many particle Hamiltonian is bounded from below allows
for a ground state.

\subsubsection{Quasi-free representations II. Reducible case}
\label{s2.3.4}
A more general class of representations is obtained by replacing
$P_-$ by a self-adjoint operator $A$ on $\cH$ with
$0<A<1$.  These representations, denoted
$\pi_A$, are constructed as follows.
We let $\cK=\cH\oplus \cH$
and form the fermion algebra
over $\cK$, denoted $\cal{A}(\cK)$. Define a projection on $\cK$ by
$$
 P(A)=\left(
\begin{array}{cc}
A & A^{1/2}(1-A)^{1/2} \\
A^{1/2}(1-A)^{1/2} & 1-A
\end{array}\right)
$$
Then the representation $\pi_A$ is
by definition the restriction of the representation
$\pi_{P(A)}$ of ${\cal{A}}(\cK)$ to the subalgebra
${\cal{A}}(\cH\oplus(0))$.

\remark These representations can be used to describe quantum field theory
models at {\em finite temperature}: If $D$ is the 1-particle Hamiltonian then
\eq
A = \frac{1}{e^{\beta D} +1}
\eqend
gives rise to the representation at temperature $T= 1/\beta>0$.
In the zero temperature limit one recovers an irreducible
representation.

\subsection{Loop groups and the boson-fermion correspondence}
\label{s2.4}
We now are ready to describe the relation
between loop groups and fermion
quantization. As underlying Hilbert space for the fermions we take
$\cH=L^2(S_L^1)\cong \ell^2(\Lambda^*_0)$ where
$$
\Lambda^*_0=\left.\left\{k=\frac{2\pi}{L}(n+\frac{1}{2})\quad\right|  \quad n\in
\Z  \right\}.
$$ These are identified via the Fourier transform,
\eq
\label{FT}
\hat f(k) =
\frac{1}{\sqrt{2\pi}}\int_{-L/2}^{L/2} dx f(x) e^{-ikx}
\eqend
for $k\in
\Lambda^*_0$.  An orthogonal basis of $L^2(S_L^1)$ is provided by the
functions
\eq
\label{ek}
e_k(x)= \frac{1}{\sqrt{2\pi}}e^{ikx}, \quad k\in \Lambda^*_0 ,
\eqend
and then we have $f=\frac{2\pi}{L} \sum_k\hat f(k)e_k.$ The spectral
projection $P_-$ we use is defined as $\widehat{(P_-f)}(k)=\hat f(k)$ for
$k<0$ and $=0$ otherwise. \\
\remark Note that $P_- = \theta(- D)$ where $D$ is the self-adjoint operator
given by $D e_k = k e_k $ for all $k\in\Lambda_0^*$.
  Of course
$D$ is a self-adjoint extension of $-i\partial_x$,
the (chiral) Dirac operator on the circle $S_L^1$.\\

Each smooth function $\alpha \in \Map(S_L^1,\C)$ naturally defines a bounded
operator on $L^2(S_L^1)$ which we denote by the same symbol,
$(\alpha f)(x)=\alpha(x)f(x)$ for all $f\in L^2(S_L^1)$. A central result in
the theory of loop groups is that all these operators $\alpha$ are in
$g_1$, and \cite{CR}
\eq
\hat S(\alpha_1,\alpha_2)= \frac{1}{4\pi}\int_{-L/2}^{L/2}
dx \left(\frac{d\alpha_1(x)}{dx}\alpha_2(x) -\alpha_1(x)
\frac{d\alpha_2(x)}{dx}\right).
\eqend
Moreover,
\eq
\label{hw}
d\Gamma(\alpha^-)\Omega=d\Gamma(\alpha^+)^*\Omega=0
\eqend
follows from Eqs.\ \Ref{cc2} and \Ref{ccc}.
Especially, all $\phi\in\cG$ are in $G_1$, and $U\to \Gamma(U)$ is
precisely the wedge-representation of $\cG$ discussed above.  The choice of
phase of $\Gamma(\phi)$ is important to obtain the explicit form for
$\sigma$ given in Eq.\ \Ref{sigma}.  To fix the phase completely we
need $R=\Gamma(\phi_1)$ corresponding to $\phi_1(x)=e^{2\pi ix/L}$ (for an
explicit construction of $\Gamma(\phi_1)$ see e.g.\ \cite{Ruij}).
This unitary operator obeys
\eq
\label{RQ}
R^{-w} d\Gamma(\alpha^\pm ) R^w = d\Gamma(\alpha^\pm), \quad
R^{-w} Q R^w = Q + w I
\eqend
for all integer $w$. Here we introduced the operator
\eq
\label{Q}
Q:=d\Gamma(I)
\eqend
which can be interpreted as the {\em charge operator}.  Writing general
loops as in Eqs.\ \Ref{f}, \Ref{alpha} we now can define
\eq
\Gamma( e^{if} ) := e^{i \bar\alpha Q/2} R^{w} e^{i \bar\alpha Q/2}
e^{id\Gamma(\alpha^+ +\alpha^-)} .
\eqend
Then a straightforward computation gives\footnote{to show this one can use
$e^{a_1} e^{a_2} = e^{[a_1,a_2]/2}e^{a_1+a_2}$ for $a_j=i
d\Gamma(\alpha_j)$, and $R^ne^{irQ} R^{-n} = e^{-inr} e^{irQ}$ for real $r$ and
integer $n$.}
\eq
\label{S1}
S(f_1,f_2)= (w_{1}\bar\alpha_2 - \bar\alpha_1 w_{2} ) + \hat
S(\alpha_1,\alpha_2) \eqend identical with Eq.\ \Ref{S}. A similar
computation implies Eq.\ \Ref{delta}.

We find it convenient to introduce normal ordering $\;\xxa \cdots \xxe$
for implementers of loops
\eq
\label{xx1}
\Gamma(e^{if}) = e^{- i S(\alpha^-,\alpha^+)/2}
\xxa\Gamma(e^{if})\xxe
\eqend
with the numerical factor chosen such that
$$\langle\Omega,\xxa\Gamma(e^{if})\xxe\Omega\rangle=1\quad \mbox{ if $w=0$}
$$ [cf.\ Eq.\ \Ref{delta}].
Note that
\eq
\label{formal}
\xxa\Gamma(e^{if})\xxe = e^{i \bar\alpha Q/2} R^{w} e^{i \bar\alpha Q/2}
e^{id\Gamma( \alpha^+)} e^{id\Gamma(\alpha^-)}
\eqend
where $e^{id\Gamma(\alpha^\pm )}$ are not operators, however,
but have to interpreted as sesquilinear forms.
This definition naturally extends to products
of implementers,
\eq
\label{xx2}
\xxa \Gamma(e^{if_1}) \Gamma(e^{if_2}) \cdots \Gamma(e^{if_N})\xxe \; := \;
\xxa\Gamma(e^{if_1}e^{if_2}\cdots e^{if_N}) \xxe
\eqend
and operators of the form
\eq
\label{normal ordering}
\xxa d\Gamma(\alpha_1)\cdots d\Gamma(\alpha_m) \Gamma(e^{f}) \xxe\; := \left.
\frac{\partial^m}{\partial a_1\cdots \partial a_m}
\xxa e^{ia_1 d\Gamma(\alpha_1)}\cdots e^{ia_m d\Gamma(\alpha_m)}
\Gamma(e^{f})
 \xxe  \right|_{a_j=0}  .
\eqend
Note that operators between normal ordering symbols commute.
We also note the following relations
\eq
\label{noo}
\xxa \Gamma(e^{if_1})\xxe \xxa \Gamma(e^{if_1})\xxe \; =
e^{-i\tilde S(f_1,f_2) }
\xxa \Gamma(e^{if_1})\Gamma(e^{if_1})\xxe
\eqend
with
\eq
\label{tildeS}
\tilde S(f_1,f_2) = w_1\bar\alpha_2 - \bar\alpha_1 w_2
+2S(\alpha_1^-,\alpha_2^+) = -S(f_2,f_1)^*
\eqend
which will be useful in the following.

\subsubsection{Bosons from fermions}
\label{s2.5}
We define $ \epsilon_p(x)= e^{-ipx}$ for $p\in  \Lambda^*$
and set
\eqa
\b(p)=d\Gamma(\epsilon_p).
\eqaend
Then $\b(-p)=\b(p)^*$, $\b(0)=Q$, and the equations given above imply
\eq
[\b(p),\b(p')] = p\frac{L}{2\pi}\delta_{-p,p'}
\eqend
and
\eq
\b(p)\Omega=0\quad p\geq 0 .
\eqend
The $\b(p)$ can be naturally interpreted as {\em boson field operators}.

\subsubsection{Fermions from bosons}
\label{s2.6}
In Ref.\ \cite{Seg1} a so-called `blip' function was
introduced which equals, up to the sign,
$$
\frac{ e^{i(x-y)2\pi/L} -\lambda }{ 1- \lambda e^{i(x-y)2\pi/L}} , \quad
0<\lambda<1 .
$$
This is the exponential of a smoothed out step
function:  Writing it as $e^{ i f_{y,\eps} }$ with $\lambda=e^{-
2\pi\eps / L }$ one gets
\eq
\label{a1}
f_{y,\eps}(x) = \frac{2\pi}{L}(x-y) +
\alpha^+_{y,\eps}(x) +\alpha^-_{y,\eps}(x)
\eqend
with
\eq
\label{a2}
\alpha^\pm_{y,\eps}(x) = \pm i \log(1- e^{2\pi (\pm i(x-y)-\eps)/L } )
=\pm i\sum_{n=1}^\infty \frac{1}{n} e^{\pm 2i\pi n(x-y)/L} e^{-2\pi\eps n/L}.
\eqend
Note that the winding number of $f_{y,\eps}$ equals $1$.
Since $f_{y,\eps}(x)$ for $\eps\downarrow 0$ converges to $ i \pi
\sgn(x-y)$ we will also use the following suggestive notation,
\eq
\label{sgn}
\sgn_\eps(x-y) := \frac{1}{\pi} f_{y,\eps}(x) .
\eqend
Later we will also need the function
$\delta_{y,\eps}= \partial_y f_{y,\eps}/2\pi$ i.e.\
\eq
\label{delta1}
\delta_{y,\eps}(x)= \frac{1}{L} + \delta^+_{y,\eps}(x) + \delta^-_{y,\eps}(x)
\eqend
with
\eq
\label{deltapm}
\delta^\pm_{y,\eps}(x)
=\frac{1}{L}\sum_{n>0} e^{\pm  2\pi i (x-y)n/L} e^{-2\pi\eps n /L }.
\eqend
This smoothed out $\delta$-function will play an important role in
Sections \ref{s4} and \ref{s5.2}.

These functions have the following important
properties\footnote{The proof is a straightforward calculation}
\eqa
\label{blips}
S(\alpha^-_{y,\eps},\alpha^+_{y',\eps'}) &=&
\alpha^+_{y',\eps+\eps'}(y) \nonu
S(f_{y,\eps},f_{y',\eps'}) &=&  \pi \sgn_{ \eps+\eps'}(y-y') \\
S(\delta^\mp_{y,\eps},\alpha^\pm_{y',\eps'}) &=&
-\delta^\pm_{y',\eps+\eps'}(y)  \nonumber
\eqaend
Note that for $\eps>0$ the operators
\eq
\label{fermion}
\phi^{\pm 1}_\eps(y)\, := \, \xxa \Gamma(e^{\pm i f_{y,\eps} }) \xxe
=\phi^{\mp 1}_\eps(y)^*
\eqend
are well-defined, and from Eqs.\ \Ref{blips} and \Ref{sigma} we
conclude
\eq
\label{exc}
\phi^{\nu}_\eps(y)\phi^{\nu'}_{\eps'}(y') = e^{i \pi
\sgn_{\eps+\eps'}(y-y')  \nu \nu'  }
\phi^{\nu'}_{\eps'}(y')\phi^{\nu}_\eps(y)
\eqend
for $\nu,\nu'=\pm 1$.  In the limit $\eps,\eps'\downarrow 0$ these formally
become anticommutator relations.  This suggests that $\phi^{\pm 1}_\eps(x)$
in the limit $\eps\downarrow 0$ should be proportional to fermion fields.
Indeed one can prove the

\vspace*{0.45cm}
{\bf \noindent  Theorem:}
\label{thm2.1}
{\it For all $f\in L^2(S_L^1)$ such that $\hat f(p)$ has a compact support
the following identity holds,
\eq
\label{bosonfermion}
\psi^*(f)=\li\frac{1}{ \sqrt{ L} }
\int_{-L/2}^{L/2} dx \,  f(x) \phi^{1}_{\eps}(y)
\eqend
in the sense of strong convergence on a dense domain.}

\vspace*{0.45cm}

This is the central result of what is usually called the
{\em  boson-fermion correspondence}; see e.g.\ \cite{CHu,F,PS}.

We now sketch a proof of this result to introduce some
techniques which we will generalize later (\cite{CR} gives a different
proof).
Note that there are subtleties which require
careful specification of the domains on which
the identities hold however we will ignore
those for brevity. The idea is to
first show by explicit computations that the
\eq
\phi^{1}(f)=\li\frac{1}{ \sqrt{ L} }
\int_{-L/2}^{L/2} dx \,  f(x) \phi^{1}_{\eps}(y), \quad
\phi^{-1}(f)=\li\frac{1}{ \sqrt{ L} }
\int_{-L/2}^{L/2} dx \,  \overline{f(x)}
\phi^{-1}_{\eps}(y) = \phi^{1}(f)^*  \quad
\eqend
obey the same CAR as the operators $\psi^{(*)}(f)$. For that we
use Eqs.\ \Ref{xx1}, \Ref{xx2}, and \Ref{blips} which imply
\eq
\label{nmo}
\phi_{\eps}^{\nu}(x)
\phi_{\eps'}^{\nu'}(y) =
b_{\teps}(r)^{\nu\nu'} \xxa \phi_{\eps}^{\nu}(x)
\phi_{\eps'}^{\nu'}(y) \xxe
\eqend
where $b_\eps(r)=-2i e^{-\pi\eps/L}\sin\PiL(r + i\eps)$, $r=x-y$, and
$\teps=\eps+\eps'$.
Thus
$$
\phi_{\eps}^{1}(x)\phi_{\eps'}^{1}(y) + \phi_{\eps'}^{1}(y)\phi_{\eps}^{1}(x)
=
-2ie^{-\pi\teps/L}\left[ \sin\PiL(r + i\teps) - \sin\PiL(r - i\teps) \right]
\xxa \phi_{\eps}^{1}(x)\phi_{\eps'}^{1}(y) \xxe
$$
where the r.h.s.\ obviously becomes zero after smearing with appropriate
test functions and sending $\eps,\eps'$ to zero. This proves
$\{\phi^{1}(f),\phi^{1}(g)\}=0$. Similarly,
$$
\phi_{\eps}^{-1}(x)\phi_{\eps'}^{1}(y) + \phi_{\eps'}^{1}(y)\phi_{\eps}^{-1}(x)
=
[\cdots]
\xxa \phi_{\eps}^{-1}(x)\phi_{\eps'}^{1}(y) \xxe
$$
where
$$
[\cdots] = \left[ \f{e^{i\pi r/L }}{1-e^{i \pi r/L }e^{-2\pi\teps/L} } +
\f{e^{-i\pi r/L }}{1-e^{-i 2\pi r/L }e^{-2\pi\teps/L} } \right] = L \delta_{y,\teps}(x)
$$
with the smoothed out $\delta$-function introduced above. Since
$\xxa \phi_{\eps}^{-1}(x)\phi_{\eps'}^{1}(y) \xxe$
becomes the identity operator for $x=y$ and $\eps=\eps'$, a simple argument
implies $\{\phi^{1}(f), \phi^{-1}(g)\}=(f,g)I$, and this completes the
proof of the CAR. Next one shows
\eq
\label{fstscd}
\phi^{-1}(P_+f)\Omega=0=\phi^1(P_-f)\Omega  \: .
\eqend
To prove this
we use Eqs.\ \Ref{fermion}, \Ref{formal}, \Ref{hw} and \Ref{RQ}
which imply
$$
\phi^{\pm 1}_\eps(x) \Omega = e^{- i \pi x /L }
e^{\pm id\Gamma( \alpha^+_{x,\eps})} R^\pm \Omega \: .
$$
Now
$$
d\Gamma(\alpha^+_{x,\eps}) = \f{2\pi i}{L} \sum_{p>0} \f{1}{p} e^{- ip x-|p|\eps }\b(-p)
$$
($p\in\Lambda^*$; cf.\ Eqs.\ \Ref{a1}--\Ref{a2}), i.e.\ it is a sum of terms with
negative Fourier coefficients only. By expanding the second exponential on the
r.h.s.\ of this equation one can therefore only generate terms with negative
Fourier coefficients, and thus $\int_{-L/2}^{L/2} dx \, e^{ikx} \psi^{\pm 1}_\eps(x) = 0$
for all negative $k\in\Lambda_0^*$. This proves Eq.\ \Ref{fstscd}.

These arguments show that
the operators $\phi^{\pm 1}(f)$ also provide the same
quasi-free representation of
the CAR on $\cF$ as the $\psi^{(*)}(f)$,
and with the same vacuum vector $\Omega$.

To complete the
proof one can check by explicit computation that the operators
$\phi^{\pm 1}(f)$ have the same commutator resp.\ exchange relations
with the operators $\b(p)$ resp.\ $R$ as the $\psi^{(*)}(f)$
and use the following

\vspace*{0.45cm}

{\bf \noindent  Lemma:}
\label{thm22}
\cite{CHOU} {\it The vectors
\eq
\prod_{n=1}^\infty \b(-\zPiL n)^{m_n} R^\nu\Omega
\eqend
with $m_n\in\N_0$ and $\nu\in\Z$ such that $\sum_{n=0}^\infty m_n<\infty$,
are dense in $\cF$.
}

\section{Quantum field theory in 1+1 dimensions}
\label{s3}
In this Section we discuss quantum field theory models of interacting
fermions on one dimensional space.  To be specific we concentrate on the
Luttinger model \cite{LM}, a simple model for a one dimensional metal.  Our
first purpose is to illustrate how the mathematical results summarized
above are used to construct and solve 1+1 dimensional quantum field theory
models.  Our second purpose is to give a physical motivation for various
operators which we construct and study in the next Section.

\subsection{The Luttinger model}
\label{s3.1}
We start with a physical motivation for this model.  We consider spinless
fermions in a one dimensional metal (wire) of length $L$ which can be
characterized by a band relation $E(p)=E(-p)$ describing the energy as a
function of the (pseudo-) momentum $p$.  If the band is filled up to the
chemical potential $\mu$, the Fermi surface consists of two points $p=\pm
p_F$ where $E(p)-\mu$ vanishes.  Physically one expects that the states
close to the Fermi surface are the most important ones.  For those one can
Taylor expand the band relations about the Fermi surface, and one gets two
branches,
\eq
\label{Taylor}
E(\pm p_F \pm(k-p_F))-\mu = \pm v_F(k- p_F) +\half m^{-1}(k -p_F)^2 +\ldots
\eqend
where $v_F$ (Fermi velocity) is the slope and $m^{-1}$ (inverse mass) the
curvature of the band at the Fermi surface. With that we obtain a
multi particle Hamiltonian $H_0=v_F (W^2_+ + W^2_-) +\half m^{-1}(W^3_+ +
W^3_-)+\ldots $ where
\eq
\label{Qs}
W^{s+1}_\pm = \int_{-L/2}^{L_2} dx \, \psi_\pm^*(x)(\pm \hat
p)^s\psi_\pm(x), \quad \hat p=- i\frac{d}{dx}
\eqend
for $s=1,2$ with $\psi_\pm$ the fermion field operators describing the
excitations of the two branches.  The model $H_0$ describes non-interacting
fermions and thus trivially is soluble.  However, if one only takes into
account the linear term in the Taylor expansion Eq.\ \Ref{Taylor}, the
model remains soluble even in presence of an interaction.

The Luttinger model thus is formally defined by the
Hamiltonian $H=H_0+H'$ where\footnote{we set $v_F=1$}
\eq
H_0=\int_{-L/2}^{L/2} dx\;
\psi^*(x) \sigma_3 \hat p \psi(x),\quad
\psi^*=(\psi_+,\psi_-)
\eqend
[$(\sigma_3)_{\sigma\sigma'}=\sigma\delta_{\sigma,\sigma'}$] is the free
part, and
\eq
H' = \int_{-L/2}^{L/2} dx\int_{-L/2}^{L/2} dy\;
\rho_+(x)v(x-y)\rho_-(y),\quad \rho_\pm(x)=\psi_\pm^*(x)\psi_\pm(x)
\eqend
the interaction (the interaction potential $v$ will be further specified
below).  It is worth noting that $H_0$ equals a Hamiltonian of free
relativistic fermions in 1+1 dimensions.  A crucial point in the correct
treatment of the model is the construction of the fermion fields $\psi_\pm$
\cite{LM}: if one would use `naive' fermions with a `vacuum' $\Omega_{\rm
unphys.}$ such that $\psi_{\pm}\Omega_{\rm unphys.}=0$, the Hamiltonians
$H_0$ and $H$ would not be bounded from below.  Since there is no
groundstate then, this model would be unphysical.  The physical idea for
solving this problem is the `filling of the Dirac sea'.  The theory of
quasi-free representations of CAR algebras described is a general formalism
which allows one to construct physical representations of the fermion
fields for {\em non-interacting} relativistic fermion models.  It turns out
that {\em the quasi-free representation in which $H_0$ is positive is also
the one in which $H$ exists and also is bounded from below}.  This is also
the case for other 1+1 dimensional models mentioned further below.  It is
this precisely this property which makes these 1+1 dimensional models
simpler than corresponding models in higher dimensions.

We now describe how to construct the physical representation for the
free Hamiltonian $H_0$. The 1-particle Hilbert space is
$L^2(S_L^1)\otimes\C^2$, and writing functions in this space as
$f=(f_+,f_-)$ we define $\widehat{(P_-f)_\pm}( \mp k)=\hat f_\pm(\mp
k)$ for all $k>0$ and $=0$ otherwise ($k\in\Lambda_0^*$; $\hat f$ as
in Eq.\ \Ref{FT}).  Then $\pi_{P_-}$ is the physical representation of
the CAR algebra $\cA$ over $L^2(S_L^1)\otimes\C^2$: The operator
$H_0=d\Gamma(-i\sigma_3 d/dx)$ is self-adjoint and positive.
Moreover, $\b_\pm (p) = d\Gamma(\half (1\pm \sigma_3)\epsilon_p )$,
$\epsilon_p(x)=e^{-ipx}$, can be identified with the Fourier modes of
the fermion currents $\rho_{\pm}(x)$.  Thus \eq H'=\frac{2\pi}{L}
\sum_{p\in\Lambda^*} \b_+(p)\hat v(p)\b_-(-p) \eqend which can be
shown to be such that $H=H_0+H'$ is self-adjoint and bounded from
below if and only if the following condition holds \cite{LM}, \eq
\label{HS}
|\hat v(p)|<1, \quad \sum_{p\in\Lambda^*} |p||\hat v(p)|^2<\infty
\eqend
where $\hat v(p)=\frac{1}{2\pi} \int_{-L/2}^{L/2} dx\, v(x) e^{-ipx}$ are
the Fourier modes of the interaction potential.  To complete the
construction of the model, one can specify a common dense invariant domains
of definition for all operators $H_0$, $H'$, $\b_{\pm}(p)$ etc.,
see e.g.\ \cite{GL2}.

We note that the results described in Subsection~2.6{\em ff} immediately
apply to the fermions $\psi_+$.  It is clear that there are similar
formulas for the $\psi_-$-fermions.  Especially, due to the non-trivial
representation $\pi_{P_-}$, the commutators of the fermion
currents are not zero, but equal to Schwinger terms.  This allows
the interpretation of the fermion currents as boson fields, as discussed.
The appearance of this Schwinger term in the commutator relations of the
fermion currents is an example of an {\em anomaly}.  It has drastic
consequences for the physical properties of the model.

The important relation which allows a solution of the Luttinger model is
the so-called {\em Kronig identity}\footnote{we sketch a proof of this
relation below}
\eq
\label{Kronig}
W^2_\pm = \frac{\pi}{L}\sum_{p\in\Lambda^*}
\xxa \b_\pm(p)\b_\pm(-p) \xxe .
\eqend
Physically this means that the free fermion Hamiltonian equals a free boson
Hamiltonian.  Since $H'$ is also quadratic in the boson fields, the
Luttinger Hamiltonian $H$ equals a free boson Hamiltonian which is
diagonalized by a unitary operator $\cU$ which can be constructed
explicitly \cite{LM,HSU}.  Then the ground state of the Luttinger model is
found as $\cU\Omega$.  Moreover, one can also compute all Green function of
the model explicitly.  This is due to the boson-fermion correspondence
which allows one to write the fermions $\psi_\pm(x)$ as a limit of
exponential of boson fields.  This means it is possible to compute the
`interacting fermion fields' $\Psi(t,x):=
\cU(t)^*\psi_\pm(x)\cU(t)$, $\cU(t)=e^{-itH }\cU$, explicitly.
The computation of Green functions reduces then to normal ordering of
products of implementers using Eqs.\ \Ref{noo}, \Ref{tildeS} \cite{HSU}.

The construction and solution for the Luttinger models described here was
for zero temperature.  A similar construction and solution of the Luttinger
model at finite temperature was given in \cite{CHa}.

\subsection{Other models}
\label{s3.2}
In the limit where space becomes infinite, $L\to\infty$, and
the interaction local, i.e.\
\eq
\label{loc}
\hat v(p)= g \quad \mbox{   independent of $p$}, \quad |g|<1,
\eqend
the Luttinger model reduces to the {\em massless Thirring model} \cite{Th}.
This latter limit is non-trivial and quite instructive: for the potential
Eq.\ \Ref{loc} the condition \Ref{HS} fails, and the operator $\cU$ does
not exist.  To construct this limit, one needs an additional multiplicative
regularization.  Due to this, the interacting fields $\Psi(t,x)$ for the
massless Thirring model are not fermions but more singular (this is nicely
explained in \cite{W}, e.g.).  To see in detail how the interacting fields
turn from fermions to these more singular operators, one can construct the
Thirring model as a limit $\ell\to 0$ of the Luttinger models with
potentials\footnote{this specific form of the `regularized' local
interacting potential results in simple explicit formulas for the
interacting fields} $
\hat v_\ell(p) = g (g^2+(1-g^2)e^{\ell |p| } )^{-1/2}
$
\cite{GLR} (for an alternative approach see \cite{CRW}).

Other interacting quantum field theory models which can be constructed and
solved by similar methods include the {\em Schwinger model} \cite{M},
i.e.\ 1+1 dimensional quantum electrodynamics with massless fermions, the
{\em Luttinger-Schwinger model}, i.e.\ the gauged Luttinger model
\cite{GLR}, and {\em diagonal QCD$_{1+1}$} \cite{CW}.  A similar
construction of {\em QCD$_{1+1}$}, i.e.\ the non-abelian version
of the Schwinger model, was given in \cite{LS}.

\section{$W_{1+\infty}$-algebra: Generalizing the Kronig identity}
\label{s4}

We now discuss an interesting mathematical application of
the formalism in Section~\ref{s2} to the so-called
$W_{1+\infty}$-algebra (see e.g.\ \cite{KR0}).

As motivation, we recall from the last Section that one can
interpret the operators $H_0 = v_F W^{2} +\half m^{-1} W^3
+\ldots $, $W^{s}=W^{s}_+$ Eq.\ \Ref{Qs}, as a (part of a)
fermion Hamiltonian.\footnote{As
in Section~\ref{s2} we only consider one branch $\psi=\psi_+$
of fermions here.} In this section we show that
the $W^s$ are examples of operators
which represent elements in the algebra
$W_{1+\infty}$, and moreover that the Kronig
identity Eq.\ \Ref{Kronig} for $W^2$ is only `the tip of
an iceberg'. There is a beautiful generalization of
the Kronig identity to the full $W_{1+\infty}$-algebra.
One purpose of our discussion here is to explain the reasoning
and methods which we will be essential in our construction of
the second quantized Calogero-Sutherland model.

\subsection{Definition of $W_{1+\infty}$}

The $W_{1+\infty}$-algebra is a central extension of
a Lie algebra $w_\infty$ defined as follows.
Consider the differential operators
\eq
\label{wsp}
w^s_p\, := \, e^{-ipx/2} (-i\partial_x)^{s-1} e^{-ipx/2}
\eqend
for  $p\in  \Lambda^*$ and $s\in\N$. It is easy to see that
these operators generate a Lie algebra with
the Lie bracket given by the commutator. To write
the commutator relations for these operators
without a lengthy
derivation it is convenient to
proceed less formally and introduce the `generating function'
\eq
\label{w1}
w_p(a) = \sum_{s=1}^\infty \frac{(-ia)^{s-1}}{(s-1)!} \, w^s_p \: ,
\quad p\in\Lambda^*
\eqend
i.e.\ $w_p(a)= e^{-ipx/2}e^{-a \partial_x} e^{-ipx/2}$,
is to be understood in the sense of formal power series in $a$.
We then compute
\nonueqa
w_p(a)w_q(b) =  e^{-ipx/2}e^{-a \partial_x} e^{-ipx/2}
 e^{-iqx/2}e^{-b \partial_x} e^{-iqx/2}
= \\
 e^{-ipx/2} e^{-iq(x-a)/2}
e^{-a \partial_x}e^{-b \partial_x} e^{-ip(x+b)/2}
 = e^{i(qa-pb)/2}w_{p+q}(a+b)
\nonueqaend
and thus obtain
\eq
\label{w2}
[ w_p(a),w_q(b) ] = (e^{i(qa-pb)/2} - e^{-i(qa-pb)/2})\, w_{p+q}(a+b) \: .
\eqend
The Lie algebra $w_\infty$ is defined by Eqs.\ \Ref{w1} and \Ref{w2}
(these relations do not really depend on the `generating function'
argument we used to write them down).
Similarly, the $W_{1+\infty}$-algebra is generated by elements $W^s_p$
collected in a `generating function'
\eq
\label{Winfty1}
W_p(a) = \sum_{s=1}^\infty \frac{(-ia)^{s-1}}{(s-1)!}\,  W^s_p \: ,
\quad p\in\Lambda^*
\: ,
\eqend
together with a central element $c$,
\eq
\label{Winfty0}
{[} W_p(a), c {]} = 0 \: ,
\eqend
and the relations
\eq
\label{Winfty}
[W_p(a),W_q(b)] = (e^{i(qa-pb)/2} - e^{-i(qa-pb)/2})\, W_{p+q}(a+b)
+
c\, \delta_{p,-q}  \frac{\sin(\frac{p}{2}(a+b)) }{ \sin(\PiL(a+b)) } \: .
\eqend

\remark One can check by direct calculation that the bracket
defined in Eq.\ \Ref{Winfty} obeys the Jacobi identity.
We will of course give a representation of this Lie algebra
next which will make it clear in what sense we interpret the
generators of this algebra as operators.
We also note that Eqs.\ \Ref{Winfty} and \Ref{Winfty1} imply
\eq
[W^1_p,W^1_q] = (p-q)W^1_{p+q} + \delta_{p,-q}\frac{L}{2\pi}
\frac{c}{12}\, p \left( p^2-\left( \frac{2\pi}{L}\right)^2 \right)
\eqend
which shows that $W^1_p$ and $c$ generate the
{\em Virasoro algebra} $Vir$:\footnote{To see that these
are indeed the usual defining relations of $Vir$,
set  $L_p\equiv W^1_p$ and $L=2\pi$ so that $\Lambda^*=\Z$.}
{\em $Vir$ is a Lie subalgebra of $W_{1+\infty}$.}

\subsection{Fermion representation of $W_{1+\infty}$}
\label{s4.1}
We can naturally identify the differential operators in Eq.\ \Ref{wsp}
with operators on the Hilbert space $L^2(S^1_L)$ defined as \eq
\label{wsp1}
w_p^s e_k = \left(k-\frac{p}{2}\right)^{s-1} e_{k-p}
\quad \forall k\in\Lambda_0^*
\eqend
for $e_k$ given by Eq.\ \Ref{ek}.
{}From our general results in Section~\ref{s2.3}
we thus expect that the operators $d\Gamma(w_p^s)$
should give a representation of a central extension of $w_\infty$
Indeed one can prove the

\vspace*{0.45cm}

{\bf \noindent  Theorem:}
\label{thm4.1}
{\it The operators
\eq
\label{wsp2}
d\Gamma(w_p^s)
\eqend
with $w_p^s$ as in Eq.\ \Ref{wsp1}, and
$c\equiv I$, give a unitary highest weight
representation of $W_{1+\infty}$, i.e.\ the relations
in Eqs. \Ref{Winfty0}--\Ref{Winfty1} and in addition,
using the notation $d\Gamma(w_p^s)\equiv W^s_p$, we have
\eq
\label{ww1}
( W^s_p )^* =  W^s_{-p} \quad \forall p
\eqend
and
\eq
\label{ww2}
W^s_p \, \Omega = 0 \quad \forall p\geq 0
\eqend
hold true for all $s\in\N$ on some common, dense, invariant domain.}

\vspace*{0.45cm}
The use of the notation $W^s_p$ to denote the
generators of $W_{1+\infty}$  in this particular representation
will not cause any confusion as no other representations
are introduced here.
To prove this theorem  one only needs to show that all
$w_p^s \in g_1$ so that the general results in Section
\ref{s2} apply. In particular, the relations in
Eq.\ \Ref{Winfty} follow from Eqs.\ \Ref{cc1}--\Ref{cc12}
where the central term is obtained from
\nonueqa
i S(w_p(a),w_q(b))
 = \sum_{k\in\Lambda_0^*}
<e_k, ( P_-w_p(a)P_+w_q(b)P_- - P_-w_q(b)P_+w_p(a)P_-) e_k>
\nonueqaend
by a straightforward computation
(use $w_p(a)e_k = e^{-ia(k-\frac{p}{2})}e_{k-p}$,
$P_\pm e_k = \theta(\pm k)e_k$ and
$<e_k,e_{k'}> = \delta_{k,k'}$). Moreover, Eq.\ \Ref{ww1} follows
from Eq.\ \Ref{ccc} and  $(w_p^s)^*=w_{-p}^s$ (the
latter can be easily checked using the definition Eq.\ \Ref{wsp1}),
and Eq.\ \Ref{ww2} follows from Eq.\ \Ref{cc2}.

We also note
\eq
\label{ww3}
[W_p^s,\hat\psi^*(k)] = \left(k-\frac{p}{2}\right)^{s-1}
\hat\psi^*(k-p)
\eqend
which follows from Eqs.\ \Ref{cc1} and  \Ref{wsp1}.

\subsection{Boson representation of $W_{1+\infty}$}
\label{s4.2}
We recall the Kronig identities which played a central role for solving the
Luttinger model, $W_0^2=\frac{\pi}{L}
\sum_{p\in\Lambda^*}\xxa\b(p)\b(-p) \xxe$.
It is well-known that this identity has a
generalization to the Virasoro algebra i.e.\ all operators
$W_p^2$ (this is the Sugawara construction; see e.g.\
\cite{GO}). We now  ask: Is there a generalization of the Kronig identity
to the full $W_{1+\infty}$-algebra?

The desired result is summarized in the following

\vspace*{0.45cm}

{\bf \noindent  Theorem:}
\label{thm4.2}
{\it Let
\eq
\label{cWy}
\cW_{\eps}(y;a) =
N(a) \left( \xxa \phi_\eps^1(y+\frac{a}{2})
\phi_\eps^{-1}(y-\frac{a}{2}) \xxe -I \right)  ,
\eqend
with the normalization constant
\eq
N(a) =\frac{i}{ 2 L \sin(\PiL a) }.
\eqend
Then the operators defined by the following equation
\eq
\label{res1}
\cW_p(a) := \li\int_{-L/2}^{L/2} dy\,
e^{-ip y }
\cW_{\eps}(y;a) =
\sum_{s=1}^\infty \frac{ (-i a)^{s-1} }{(s-1)!}\,  \cW^{s}_p
\eqend
equal the operators in Eq.\ \Ref{wsp2}: $\cW_p^{s}=W_p^{s}$ for all
$s\in\N$ and $p\in\Lambda^*$.}

\vspace*{0.45cm}

To see that this theorem allows us to  compute formulas for all the operators
$W^s_p$ in terms of the boson operators $\b(p)$ we write
$\;\xxa
\phi_\eps^1(y+\frac{a}{2})\phi_\eps^{-1}(y-\frac{a}{2})
\xxe = \; \xxa e^{i\nu[\cdots]} \xxe\;\;$
with
$$
[\cdots]= d\Gamma(f_{y+\frac{a}{2},\eps}) -
d\Gamma(f_{y-\frac{a}{2},\eps}) =
-2\pi \left[ a \rho_\eps(y) +
\frac{a^3}{24} \partial_y^2 \rho_\eps(y) +
\ldots \right]
$$
where $\partial_y=\partial/\partial y$, and
\eq
\rho_\eps(y)=d\Gamma(\delta_{y,\eps}) = \frac{1}{L}\sum_{p\in\Lambda}
\b(p)e^{ipy}e^{-|p|\eps}
\eqend
is the regularized  fermion current in position space.  Inserting this
in the l.h.s. of Eq.\ \Ref{res1}, expanding in powers of $a$ and comparing
with the r.h.s.  of Eq.\ \Ref{res1} one obtains
\eqa
\label{res2}
W^{1}_p &=&  \nt e^{-ipy} \xxa \rho_\eps(y)\xxez = \b(p) \nonu
W^{2}_p &=&  \pi \nt  e^{-ipy} \xxa \rho_\eps(y)^2 \xxez =
\frac{1}{2}\left(\frac{2\pi}{L}\right) \sum_{q\in\Lambda^*}
\xxa \b(q)\b(p-q) \xxe
\nonu W^{3}_p &=&\frac{4\pi^2}{3}\nt  e^{-ipy}  \xxa \left(
\rho_\eps(y)^3 -\frac{1}{4L^2}\rho_\eps(y) \right) \xxez
\\
&=& \frac{1}{3}\left(\frac{2\pi}{L}\right)^2\sum_{q_1,q_2\in\Lambda^*}
  \xxa \b(q_1)\b(q_2)\b(p-q_1-q_2)\xxe +
\ldots
\nonu
&  & \vdots  \nonu
W^{s+1}_p &=& \frac{1}{s+1}\left(\frac{2\pi}{L}\right)^{s}
\sum_{q_1,\ldots, q_{s}\in\Lambda^*}
  \xxa \b(q_1)\cdots \b(q_{s})\b(p-q_1-\cdots - q_{s})\xxe +\ldots
\nonumber
\eqaend
where `$+ \ldots$' refers to those
terms involving fewer $\b$'s.

We now sketch how this theorem can be proved by using the results
summarized in Section~\ref{s2}.  We recall Eq.\ \Ref{bosonfermion}
which shows that $L^{-1/2}\phi^1_\eps(y)=L^{-1/2}\xxa\Gamma(e^{i
f_{y,\eps} })\xxe$ equals a regularized fermion operator $\psi^*(y)$.
Using that the argument is simple: we compute the commutator of
$\cW_{\eps'}(y;a)=N(a)\xxa\Gamma(e^{i[ f_{y+\frac{a}{2},\eps'}-
f_{y-\frac{a}{2},\eps'} ]} )\xxe$ with $\phi^1_\eps(x)$ using
Eqs.\ \Ref{noo}, \Ref{tildeS} and \Ref{blips}. We obtain
$$ [\cW_{\eps'}(y;a), \phi^1_\eps(x)] = (\cdots)
\xxa\Gamma(e^{i[f_{x,\eps}+ f_{y+\frac{a}{2},\eps'}-
f_{y-\frac{a}{2},\eps'} ]})\xxe $$
with\footnote{at this point the reason for our normalization constant
$N^1(a)$ becomes obvious.}
$$ (\cdots)\, =
N(a)\left( \frac{\sin\PiL(y+\frac{a}{2}-x+i\tilde\eps) }
{\sin\PiL(y- \frac{a}{2}-x+i\tilde\eps) }
-c.c.\right) = \frac{i}{2 L}\left(
\cot\PiL(y- \frac{a}{2}-x+i\tilde\eps)  - c.c. \right)
$$
where $\tilde\eps=\eps+\eps'$ and $c.c.$ means the same terms complex
conjugated. We now observe that\footnote{This is easily seen by expanding
the l.h.s as a Taylor series in $e^{\pm i(y-x)2\pi/L}e^{-\eps 2\pi/L}$.}
\eq
\label{ctg}
\pm\frac{i}{2 L} \cot\PiL(y-\frac{a}{2}-x\pm i\tilde\eps) = \frac{1}{2L} +
\delta^\pm_{x,\tilde\eps}(y-\frac{a}{2}),
\eqend
which implies $(\cdots)=\delta_{x,\tilde\eps}(y-\frac{a}{2})$. We
conclude that
$$
\lim_{\eps'\downarrow 0}\int_{-L/2}^{L/2} dy\, e^{-ipy}
[\cW_{\eps'}(y;a), \phi^1_\eps(x)] \, =\, e^{-ip(x+\frac{a}{2}) }
\xxa\Gamma(e^{if_{ x+a ,\eps}})\xxe  $$
equivalent with
$$
[\cW(a), \phi^1_\eps(x) ] = \, = \,
e^{-ip(x+\frac{a}{2}) } \phi^1_\eps(x+a)
$$
Using now
$$
\hat \psi^*(k)=\li\frac{1}{ L}
\int_{-L/2}^{L/2} dx \,  e^{ikx} \phi^{1}_{\eps}(y)
$$
which follows from Eq.\ \Ref{bosonfermion} we conclude that
$$
[\cW_p(a), \hat\psi^*(k)] = e^{-ia(k-\frac{p}{2})}
\hat\psi^*(k-p) \: .
$$
Recalling Eq.\ \Ref{res1} and \Ref{ww3} we thus see that
$[\cW_p^s, \psi^*(k)]=[W_p^s ,\psi^*(k)]$ always.
Moreover, by definition of $\;\xxe\cdots\xxe$ we also
get $\cW(a)_p\Omega=0$ for all $p\geq 0$, i.e.\
$\cW^{s+1}_p\Omega = 0$ for all $p\geq 0$.
It is also easy to check that
$(\cW_p^s)^* = \cW_{-p}^s$, and the theorem therefore
follows by applying the following Lemma to
$A=\cW_p^{s}-W^{s}_p$ for $p\geq 0$.

\vspace*{0.45cm}

{\bf \noindent  Lemma:}
\label{thm4.3} \cite{CR} {\it For linear operators $A$ on $\cF$,
$[A,\hat\psi^*(k)]=0$ for all $k\in\Lambda^*_0$, and $A\Omega=0$
imply $A=0$. }

\section{Anyons and the Calogero-Sutherland model}
\label{s5}
\subsection{Boson-anyon correspondence}
\label{s5.1}
In this section we discuss how to generalize the boson-fermion
correspondence to anyons.

\subsubsection{Construction of anyon field operators}
To construct {\em any}ons we have to extend the relations Eq.\ \Ref{exc} to
{\em any} non-integer $\nu\nu'$.  The naive idea would be to define
$\phi^\nu_\eps(y)=\;\xxa \Gamma(e^{i\nu f_{y,\eps}})\xxe$ for arbitrary $\nu$,
and these objects would then (formally) obey the desired relations.
However, since the functions $e^{i\nu f_{y,\eps}(x)}$ are {\em not}
periodic if $\nu$ is not an integer, the operator $\Gamma(e^{i\nu
f_{y,\eps}})$ does not exist
in the Fock representation in general.  This technical difficulty
indicates that anyon field operators are delicate objects whose consistent
construction requires some care.

To circumvent this problem, we note that $S(f_1,f_2)$ Eq.\ \Ref{S1} is
invariant under changes $\bar \alpha_i\to \bar \alpha_i \lambda$ and $w_i\to
w_i/\lambda$ with an arbitrary scaling parameter $\lambda$.  We use this to construct a
function $\tilde f_{y,\eps}(x)$ which has the following properties,
\eq
\bma{ll} (i)& \quad e^{i \nu \tilde f_{y,\eps}(x)}
\quad\mbox{ is periodic for all $\nu$,}\\
(ii)& \quad S(\tilde f_{y,\eps},\tilde f_{y,\eps})=
S(f_{y,\eps},f_{y,\eps}).\nonumber
\ema
\eqend
Since the functions $\nu \tilde f_{y,\eps}(x)$ have winding numbers
different from zero, the first requirement can only be fulfilled
for $\nu$ values which are an integer multiple of some fixed number
$\nu_0>0$.  Then
\eq
\tilde f_{y,\eps}(x) =
\frac{2\pi}{L\nu_0}x -\frac{2\pi\nu_0}{L}y
+ \alpha^+_{y,\eps}(x) +\alpha^-_{y,\eps}(x)
\eqend
has the desired properties. Thus the operators
\eq
\label{anyon}
\phi_{\eps}^{\nu}(y):=\; \xxa\Gamma(e^{ i\nu \tilde f_{y,\eps}})\xxe\;
=\phi_{\eps}^{-\nu}(y)^*, \quad \nu/\nu_0\in\Z \eqend are well-defined
for $\eps>0$, and they obey the exchange relations Eq.\ \Ref{exc} but
now for all $\nu,\nu'$ which are integer multiples of $\nu_0$.  Thus
the theory of loop groups provides a simple and rigorous construction
of regularized free anyon field operators $\phi_{\eps}^\nu(x)$.

\subsubsection{Anyon correlation functions}
\label{s5.1.2.}
We now can easily compute all anyon correlations functions:
Eqs. \Ref{xx1}, \Ref{xx2}, and \Ref{blips} imply
\eq
\label{no}
\phi_{\eps_1}^{\nu_1}(y_1)\cdots
\phi_{\eps_N}^{\nu_N}(y_N) =
\cJ^{\nu_1,\cdots,\nu_N}_{\eps_1,\cdots,\eps_N}(y_1,\ldots,y_N)
\xxa \phi_{\eps_1}^{\nu_1}(y_1) \cdots
\phi_{\eps_N}^{\nu_N}(y_N)\xxe
\eqend
where
\eq
\label{J}
\cJ^{\nu_1,\cdots,\nu_N}_{\eps_1,\cdots,\eps_N}(y_1,\ldots,y_N)
=
\prod_{j<k} b(y_j-y_k; \eps_j+\eps_k)^{\nu_j\nu_k}
\eqend
with
\eq
\label{b}
b_\eps(r) = -2i e^{-\pi\eps/L}\sin\PiL(r + i\eps).
\eqend
With Eq.\ \Ref{delta} we obtain
\eq
\label{corr}
\left<\Omega,  \phi_{\eps_1}^{\nu_1}(y_1)\cdots
\phi_{\eps_N}^{\nu_N}(y_N)\Omega\right> = \delta_{\nu_1+\ldots
+\nu_N,0}
\cJ^{\nu_1,\cdots,\nu_N}_{\eps_1,\cdots,\eps_N}(y_1,\ldots,y_N) .
\eqend

\subsection{Second quantized Calogero-Sutherland Hamiltonian}
\label{s5.2}
We recall the operators $W^{s+1}\equiv
W_{p=0}^{s+1}$ which (formally) obey (see Eq.\ \Ref{ww3})
\eq
\label{formal1}
[W_0^{s+1}, \psi^*(x) ] =
i^s\f{\partial^s}{\partial x^s} \psi^*(x)
\eqend
where $\psi^*(x)=\phi^1(x)$ are the fermion operators.
We now try to find generalizations of these operators
to the case of anyons. Without loss of generality, we assume
in the following $\nu=\nu_0>0$.

We first generalize the operators  $W^{s+1}\equiv W^{s+1}_{p=0}$
defined in Eqs.\ \Ref{cWy}--\Ref{res1} to general $\nu$,
\eq
\label{res11}
\cW^\nu(a) := \li\int_{-L/2}^{L/2} dy\,
\cW^\nu_{\eps}(y;a) =
\sum_{s=1}^\infty \frac{ (-i a)^{s-1} }{(s-1)!} W^{\nu,s}
\eqend
with
\eq
\label{cWy1}
\cW^\nu_{\eps}(y;a) : =
N^\nu(a) \left( \xxa e^{i\nu d\Gamma(\tilde f_{y+a,\eps}- \tilde
f_{y,\eps}) } \xxe - I \right)
\eqend
and
\eq
\label{Na1}
N^\nu(a) =\frac{i}{ 2 L \nu^2\cos^{\nu^2}(\PiL a)\tan(\PiL a) } .
\eqend
Similarly as in the Proof of Theorem \ref{thm4.2} we compute
\eq
\label{comm}
[\cW^\nu_{\eps'}(y;a), \phi^\nu_\eps(x)] = (\cdots)
\xxa\Gamma(e^{i\nu [\tilde f_{x,\eps}+ \tilde f_{y+a,\eps'}-
\tilde f_{y,\eps'} ]})\xxe
\eqend
with
\nonueqa (\cdots):=
N^\nu (a) \left[ \left(\frac{\sin\PiL(y+a-x+i\tilde\eps) }
{\sin\PiL(y-x+i\tilde\eps) }\right)^{\nu^2} -c.c.\right]\\ =
N^\nu(a)\cos^{\nu^2}(\PiL a)\left(1+\tanh(\PiL a)
\cot\PiL(y-x+i\tilde\eps) \right)^{\nu^2}  + c.c.
\nonueqaend
and $\tilde\eps=\eps+\eps'$. Expanding this in powers of $a$ and using
$\cot^2(z)=-1-d\cot(z)/dz$ and Eq.\ \Ref{ctg} we obtain
\eq
\label{cdots}
(\cdots)=\delta_{x,\tilde\eps}(y) -\half(\nu^2-1)a\partial_y
\delta_{x,\tilde\eps}(y) + \cO(a^2).
\eqend
Thus
$$
[\cW^\nu(a), \phi^\nu_\eps(x)] = \phi^\nu_\eps(x+a)
+ i \pi \nu(\nu^2-1)a \xxa [\tilde\rho_\eps(x+a)-\tilde\rho_\eps(x)]
\phi^\nu_\eps(x+a) \xxe +\cO(a^3) .
$$
Comparing now equal powers of $a$ on both sides of Eq.\ \Ref{comm} we
see that the generalization of Eq.\ \Ref{formal1} to anyons holds
true only for $s=0,1$,
\eq
[W^{\nu,s+1},\phi^\nu_\eps(x)] =\nu^{1-s}
i^s\frac{\partial^s}{\partial x^s}
\phi^\nu_\eps(x)\qquad  s=0,1
\eqend
but for $s>2$ we get correction terms, e.g.\
\eq
\label{w3}
[W^{\nu,3},\phi^\nu_\eps(x)] = \frac{i^2}{\nu}\frac{\partial^2}{\partial
x^2}\phi^\nu_\eps(x) + 2\pi i (\nu^2-1) \xxa
\tilde\rho_\eps(x)'\phi^\nu_\eps(x) \xxe
\eqend
where
\eq
\tilde\rho_\eps(y)=-\frac{1}{2\pi}d\Gamma(\partial_y\tilde f_{y,\eps})=\rho_\eps(y)
+\frac{\nu -1}{L}Q .
\eqend
Here and in the following we only consider the first non-trivial case
$s=2$.  We now need to cancel the second term in Eq.\ \Ref{w3}.  This can
be partly done by an operator
\eq
\cC \propto i \li\int_{-L/2}^{L/2}dy\; \xxa \rho_{y,\eps}^+
\partial_y \rho^-_{y,\eps} \xxe
\eqend
where $\rho_{y,\eps}^\pm = d\Gamma(\delta^\pm_{y,\eps})$. By explicit
computation similar to the one above one can prove that
$$
\cC \phi^\nu_\eps(x) + \phi^\nu_\eps(x)\cC =
2\pi i \xxa \tilde\rho_{2\eps}(x)'\phi^\nu_\eps(x)\xxe + 2\xxa \cC
\phi^\nu_\eps(x)\xxe .
$$
The first term can be used to cancel the second term in Eq.\ \Ref{w3}.
The second term seems somewhat strange, however, it disappears
when applying this equation to vectors of the form $R^w\Omega$, $w$ an
arbitrary integer (in contrast to the first term!).  We thus see that
the operator
\eq
\label{Hpmnu}
\cH^{\nu,3} = \nu W^{\nu,3} + (1-\nu^2) \cC
\eqend
obeys the relation
$$
[\cH^{\nu,3},\phi^\nu_\eps(x)]R^w\Omega
\simeq i^2\frac{\partial^2}{\partial x^2}
\phi^\nu_\eps(x)R^w\Omega
$$
where `$\simeq$' means `equal up to a regular term which vanished for
$\eps\downarrow 0$'.  This seems to be the best we can do to
generalize the relation Eq.\ \Ref{formal1} for $s=2$ to the anyon
case.  However, to fully appreciate this operator $\cH^{\nu,3}$, one
has to extend the computation above to a product of multiple anyon
operators \cite{CL}.  One thus obtains the following

\vspace*{0.45cm}

{\bf \noindent  Theorem:}
\label{thm5.1}
{\it There exists an operator
$\cH^{\nu,3}$ which obeys the following relations,
\eq
\label{that}
[\cH^{\nu,3}, \phi^\nu_\eps(y_1)\cdots  \phi^\nu_\eps(y_N)]R^w\Omega \simeq
H_{N,\nu^2}^\eps \phi^\nu_\eps(y_1)\cdots  \phi^\nu_\eps(y_N)R^w\Omega
\eqend
for all integer $w$, where
\eq
\label{CM}
H_{N,\nu^2}^\eps = -\sum_{k=1}^N \frac{\partial^2}{\partial y_k^2}
+\sum_{1\leq k<\ell\leq N} 2 \nu^2 (\nu^2 - 1) V_\eps(y_k-y_\ell)
\eqend
with
\eq
V_\eps(r) = -\f{\partial^2}{\partial r^2} \log b_\eps(r)
\eqend
is a regularized version of the Calogero-Sutherland Hamiltonian defined in
Eq.\ \Ref{Sutherland}. }

\subsubsection{Constructing eigenfunctions for
the Calogero-Sutherland Hamiltonian}
\label{s6.2}
We now sketch how the theorem in the previous section
can be used to find eigenfunctions of the
Calogero-Sutherland Hamiltonian: Suppose we found a common eigenvector
$\eta$ of the operators $\cH^{\nu,3}$ and $Q$,
\eq
\label{psi}
\cH^{\nu,3}\eta = E\eta, \quad Q\eta= N\eta .
\eqend
Then the theorem in Section \ref{thm5.1} and the relation
\eq
\label{not}
\cH^{\nu,3}\Omega=0, \eqend imply that \eq F_\eta (x_1,\ldots,x_N) =
\li\left<\eta, \phi_{\eps}^{\nu}(x_1) \cdots \phi_{\eps}^{\nu}(x_N)
\Omega \right>, \eqend is is an eigenfunction of the
Calogero-Sutherland Hamiltonian with the eigenvalue $E$.  This follows
immediately if we sandwich Eq.\ \Ref{that} between $\eta$ and
$\Omega$, use Eq.\ \Ref{not} and take the limit $\eps\downarrow 0$.

To find such vectors $\eta$ one again can use Eqs.\ \Ref{that} and
\Ref{not}.  The idea is to consider the Fourier modes of the anyon
field operators, \eq \hat\phi^\nu(p)=\li \int_{-L/2}^{L/2} dx\,
e^{i\pi \nu^2 Q y /L}\phi^\nu_{\eps}(x)e^{i\pi\nu^2 Q y/L} e^{i px}
\eqend where $p\in \Lambda^*$.  Note that the anyon field operators
are not periodic, and we have to remove the non-periodic factors
before Fourier transformation. Using the theorem in Section
\ref{thm5.1} one then can show that linear combinations of vectors
$$
\eta= \hat\phi^\nu(p_1)\cdots \hat\phi^\nu(p_m) R^{N-m}\Omega,
\qquad p_k-p_{k+1}\geq 0  ,\quad 0\leq m\leq N
$$ obey Eq.\ \Ref{psi} which eigenvalues $E$ one can easily compute.  The
simplest case is $\eta=R^N\Omega$ where we obtain a eigenstate which is
essentially the known groundstate of the Calogero-Sutherland model
\cite{Su} (up to a phase factor which corresponds to a non-zero center-of-mass
`motion'). In general, we obtain all the eigenfunctions of the
Calogero-Sutherland model \cite{CL} which where
found originally by Sutherland \cite{Su}.

\subsubsection{Finite temperature anyons and the elliptic
Calogero-Sutherland model} Recently the construction described above
was generalized to the elliptic Calogero-Sutherland model i.e.\ the
Hamiltonians in Eq.\ \Ref{Sutherland} with an interaction potential
\eq V(r) = -\f{\partial^2}{\partial r^2} \log\left( \sin ( \PiL
r)\prod_{n=1}^\infty [1-2q^{2n}\cos(\zPiL r)+q^{4n}] \right) \eqend
($0\leq q < 1$) which is equal, up to an additive constant, to the
Weierstrass elliptic function $\wp(r)$ with periods $L/2$ and
$i\log(1/q)$. The idea was to consider finite temperature anyons i.e.\
construct the anyons as described above but using a quasi-free
reducible fermion representation corresponding to a finite temperature
$1/\beta$, as described in Section~\ref{s2.3.4}.  The motivation came
partly from the geometric interpretation of this fermion
representation which we describe in Section~\ref{s7}.

Introducing a non-zero temperature quasi-free state
only changes the normal ordering
prescription i.e.\ all formulas in Section~\ref{s5.1.2.} remain true except
Eq.\ \Ref{b} which is changed to
\eq
\label{be}
b_\eps(r) = -2i e^{-\pi\eps/L}\sin\PiL(r + i\eps)
\prod_{n=1}^\infty [1-2q^{2n} e^{-2\pi \eps/L}\cos(\zPiL r)+q^{4n} e^{-4\pi \eps/L} ]
\eqend
where $q=\exp(-\beta L/(2\pi) )$. With that modification, the theorem in Section
\ref{thm5.1}
remains true, and this gives a second quantization
of the elliptic Calogero-Sutherland system.  To further generalize  the results
in \cite{CL} there is one complication:  for $\beta<\infty$, the operator
$\cH^{\nu,3}$ no longer obeys Eq.\ \Ref{not}, and thus the argument leading
to a solution algorithm seems to fail. However, one can prove that the weaker condition
\eq
\label{not1}
\left<\Omega, [\cH^{\nu,3},\phi_{\eps'}^{\nu}(y_N)^* \cdots \phi_{\eps'}^{\nu}(y_1)^*
\phi_{\eps}^{\nu}(x_1) \cdots \phi_{\eps}^{\nu}(x_N)]
\Omega \right> = 0
\eqend
still holds, and this is enough to obtain a solution algorithm:
the theorem in Section~\ref{thm5.1} and this relation imply
\eq
\label{id22}
H_{N,\nu^2}^\eps(\vx) F_{N,\nu^2}^{\eps',\eps}(\vy,\vx) \simeq
\overline{H_{N,\nu^2}^{\eps'}(\vy)} F_{N,\nu^2}^{\eps',\eps}(\vy,\vx)
\eqend
where $F_{N,\nu^2}^{\eps',\eps}(\vy,\vx) =
\left<\Omega,\phi_{\eps'}^{\nu}(y_N)^* \cdots \phi_{\eps'}^{\nu}(y_1)^*
\phi_{\eps}^{\nu}(x_1) \cdots \phi_{\eps}^{\nu}(x_N)
\Omega \right>$ and the Hamiltonians on the two sides act on different
arguments $\vx$ and $\vy$, as indicated. From Eqs.\ \Ref{corr} and \Ref{J}
we obtain
\eq
\label{F22}
F_{N,\nu^2}^{\eps',\eps}(\vy,\vx) =
\f{ \prod_{1\leq j<j'\leq N} b_{2\eps'}(y_{j'}-y_j)^{\nu^2}
\prod_{1\leq k<k'\leq N}
b_{2\eps}(x_{k}-x_{k'})^{\nu^2}}{\prod_{j,k=1}^N
b_{\eps+\eps'}(y_j-x_k)^{\nu^2}} \: .
\eqend
This remarkable identity Eq.\ \Ref{id22} together with Eq.\ \Ref{F22}
can be used to construct eigenfunctions
of the elliptic Calogero-Moser Hamiltonian as linear combinations
of the anyon correlation functions
$$
\li \left<\Omega, \hat \phi^{\nu}(p_N)^* \cdots \hat \phi^{\nu}(p_1)^*
\phi_{\eps}^{\nu}(x_1) \cdots \phi_{\eps}^{\nu}(x_N)
\Omega \right>
$$
which can be computed from $F^{\eps',\eps}(\vy,\vx)$ by Fourier transformation
in the variables $\vy$ and taking the limits $\eps,\eps'\downarrow 0$.
This provides a generalization of the solution of the
Calogero-Sutherland model \cite{Su} to the elliptic case
\cite{L4}.

\prt{PART B: LOOP GROUPS, 1+1 DIMENSIONAL QFT AND RIEMANN SURFACES}

\section{Overview}
\label{s6}
This part presents two main themes. First we give a
summary of some of the literature on various applications
of loop group representation theory in quantum field
theory. This list is not exhaustive and is confined to
examples in which we have had some involvement.
Second we extend the elementary discussion of the examples of the previous
section. This extension consists of two
related discussions.  First the
K.M.S. state (or finite temperature state)
on the fermion algebra over $L^2(S^1)$ for the free
Dirac Hamiltonian is shown to be interpretable geometrically as
describing fermions on a torus. This then leads into
a discussion of fermions on higher genus Riemann surfaces.
This latter exposition is more mathematically sophisticated
and assumes some knowledge of the geometry of Riemann surfaces.
The idea here is to sketch how one constructs
quantum fields and vertex operators
on Riemann surfaces from representations of loop groups.

\subsection{A guide to various examples}

We introduce some examples of quasifree representations of
the fermion field algebra following on from our exposition in 2.3.2 and 2.3.4.

\noindent{\bf Notation}
(i) We let $P_-$ denote the projection on $L^2(\R,\C^N)$ (resp.
$L^2 (S^1,\C^N)$) onto functions which are boundary values of
functions holomorphic in the lower half plane in $\C$ (resp.
exterior of the unit disc).\\
(ii) Let $A(\beta)$ denote the operator on $L^2(S^1,\C^N)$ (resp.
$L^2(\R,\C^N)$) which is given by multiplication by the
function
$$
   k\to e^{-\beta k}/(1+e^{-\beta k}),\qquad k\in \Z (\mbox{ resp.}
k\in \R)\quad (\beta\geq 0)
$$
on the Fourier transform.\\
(iii) Let $A(m)$ denote the operator on $L^2(\R,\C^N)$ given by
multiplication on the Fourier transform by the function
$$
p\to (1-p/(p^2+m^2)^{1/2})/2,\qquad (m\geq 0).
$$

These operators arise respectively as follows.

\noindent (i) The operator $P_-$
is the spectral projection of the massless Dirac Hamiltonian
corresponding to the negative part of the spectrum. Then the resulting
representation of  ${\cal{A}}$ is the usual `infinite wedge
representation' or equivalently, that obtained by `filling
the Dirac sea'.

\noindent (ii). The operator $A(\beta)$
defines a K.M.S. state
(or temperature state at inverse temperature $\beta$)
 on the Fermion algebra $\cal A$ for the one
parameter group of automorphisms generated by the massless Dirac operator
(see 2.3.4).

\noindent (iii). The operator
$P_{A(m)}$ is the spectral projection of the massive Dirac
Hamiltonian corresponding to the interval $(-\infty,-m]$.

In this exposition we cannot provide details of all
of the applications of loop groups to quantum field theory.
The following is a brief guide to
a number of papers which deal with
models in $1+1$-dimensional space-time.

$\bullet$
The standard free field construction of the basic representation
of the affine Lie algebra $A_{N-1}^{(1)}$
can be obtained by taking the underlying Hilbert space
to be $L^2(S^1,\C^N)$ and the fermion representation
to be $\pi_{P_-}$. The construction is a simple
generalization of the previous discussion of the
wedge representation of the loop group of $\U(1)$ to
the construction of the wedge representation of the loop group
 of $\U(N)$. The affine Lie algebra $A_{N-1}^{(1)}$ arises
as the Lie algebra of this central extension of the
loop group of $\U(N)$ which is acting on the fermion Fock space
for $\pi_{P_-}$.
There is also an analogous construction of
projective representations of the loop groups of  SU($N$) and SO($N$)
(see \cite{CR}).

$\bullet$ The Cayley transform from the circle to the
real line may be used to realise the preceding
affine Lie algebra representation
as the infinitesimal version of  a projective representation of the
group $\Map(\R, \U(N))$
of smooth maps from the real line into $\U(N)$.
This representation acts
on the fermion Fock space over
$L^2(\R,\C^N)$ \cite{CR}.

$\bullet$ Temperature or KMS states on $A_{N-1}^{(1)}$
may be obtained by realizing this affine algebra in the
representation space of free fermions at inverse temperature $\beta$.
Here the underlying Hilbert space is
 $L^2(S^1,\C^N)$ and the fermion representation is
$\pi_{A(\beta)}$. The corresponding state on
the fermion algebra $\cal A$ over $L^2(S^1,\C^N)$ is
a K.M.S. state. The cyclic vector (or vacuum)
for the resulting representation
of the
C$^*$-algebra generated by the operators
$$\{\rho_{A(\beta)}(\phi): \phi\in \Map(S^1, \U(N))\}$$ defines a K.M.S.
state on this algebra. Thus, in this way, we obtain
temperature states on loop groups. Generalising from
Subsection~2.6 we can use blip functions to reconstruct
the fermions in the non-zero temperature (K.M.S.)
representation $\pi_{A(\beta)}$. When $N=1$ the
correlation functions of the approximate fermion operators give
theta function identities \cite{CHa}. Interestingly
this example can be re-interpreted geometrically
as quantum field theory on the torus \cite{CH1}, a fact
which we will explain in more detail in the next section.

$\bullet$
Starting with the underlying Hilbert space
$L^2(S^1,\C)\oplus L^2(S^1,\C)$
and the fermion representation $\pi_{P_{A(\beta)}}$
we can use boson algebra automorphisms
to `twist' the vertex operators so as to obtain the
fields of the non-zero temperature Luttinger model
\cite{CHa}.

$\bullet$ By using massive fermions over
$L^2(\R,\C^N)$,
that is, the representation $\pi_{A(m)}$ of the fermion algebra
 over $L^2(\R,\C^N)$ one may obtain a type
III$_1$ factor representation of the group $\Map_0(\R, \U(N))$
consisting of smooth maps $\phi:\R\mapsto \U(N)$ with $\phi(0)=1$.
This may be used to construct sine-Gordon fields at the critical value
of the coupling constant where the theory is free \cite{CR}.

Note that in this last example if we put $m=0$ we can construct the massless
Thirring model \cite{CRW}. A less complicated version of this construction
is what we used in Subsection~3.1 for
the Luttinger model.
Finally we remark on a basic limitation of this approach.
In all the examples there is somewhere in the background
a free quantum field theory.
A situation which we
would like to understand
and for which the methods of this paper do not apply
 is the massive Thirring
model or equivalently, the sine-Gordon  model for
general coupling constant.

\section{Free fields on the torus}
\label{s7}
In this section we provide a re-interpretation of the
quasifree representation of the fermions at inverse temperature
$\beta$ as a theory of free fermions on the torus.
This is the simplest case of the more general theory
of free fields on Riemann surfaces.
We regard the torus as constructed from two annuli by joining along the
boundary.  Equivalently (at least topologically) this is the same as
joining two cylinders to form the torus.

We let $R$ be the annulus (in the complex plane)
$\{w:e^{-\beta/2}\leq|w|\leq 1\}$
and $\bar{R}$ be the annulus $\{w:1\leq|w|\leq e^{-\beta/2}\}$ where $\beta>0$.
We may now form the torus $\Sigma=R\,\sqcup\,\partial R\,\sqcup\,\bar{R}$ by
first joining $R$ and $\bar{R}$ along their common boundary and
then identifying the remaining boundary circles.
Notice that $\Sigma$ then
possesses an anticonformal involution: $w^\flat=1/\bar{w}$
(an example of the so-called Schottky involution).

With respect to the canonical homology basis
\[
   A=\{e^{\beta/2+i\theta},0\le\theta\le 2\pi\},\
   B=\{r\in{\R}:e^{-\beta/2}<r<e^{\beta/2}\},
\]
where $w=r e^{i\theta}$ are polar coordinates,
we may think of the torus as
the complex plane modulo the lattice $2\pi{\Z}+i\beta({\Z+1/2})$.

Introduce the four spin structures (square roots of the cotangent bundle)
$L^\alpha$ where $\alpha=(0,0),(1/2,0),(0,1/2),(1/2,1/2)$ denotes the
so-called theta characteristics
which summarise
the behaviour of their sections
as we shift through the periods of the $A$ and $B$
cycles.
These sections may be defined as functions on the
complex plane with well defined equivariance properties under the action of the
translations defining the lattice.  Specifically these are: periodic under both
generators, periodic under one and anti-periodic under the other and
antiperiodic under both respectively.
So for example for $\alpha =(0,1/2)$ this means the sections
regarded as functions on the complex plane satisfy
$$f(w+2\pi) = f(w) \quad f(w+i\beta)=-f(w).$$
The Hilbert spaces for the fermionic
algebras live on the boundary of $R$ (or equivalently $\bar R$)
We form pre-Hilbert spaces of
smooth sections of each $L^\alpha$ restricted to the boundary with
$L^2$ norms defined as follows.
Let $K=(L^\alpha)^2$ denote the cotangent
bundle with transition functions $z_{\gamma\delta}$ then $|K|$ is the bundle
with transition functions $|z|_{\gamma\delta}$ so that if $f$ a section of
the square root $L^\alpha$ of $K$ then $|f|^2$ is a section of $|K|$,
that is, a measure on the boundary circles. So it makes sense
to integrate $|f|^2$ on the two boundary circles to
define the norm:
$$||f||^2 = \int_{\partial R}|f|^2$$
  We denote the completions by $H^\alpha=H^\alpha(\partial R)$.
Equivalently we could work with $\bar R$
and the space $H^\alpha(\partial \bar R)$.
  The appropriate
representation of the fermion algebra ${\mathcal A}(H^\alpha)$
 is, in the case of the
 first three
spin structures,
given by the projection $P^\alpha$ onto the subspace of $H^\alpha$ which is
the closure of the subspace obtained by restriction to the boundary of the
holomorphic sections.  (For the last spin structure there is a problem
with this definition but we will not resolve it here.)

To see that this is an interesting thing to do we need to give these
projections explicitly.  In each case they are given by the Szego kernel
which in Fay's notation \cite{Fa} is written $\sigma_\alpha(\bar{x},y)$.  We
will write this down explicitly in
the case where it reproduces the
 KMS states on the loop group \cite{CHa}.

Explicitly, on restriction to the circle $\{w:|w|=1\}$
 in the boundary of $\bar R$ we have,
in polar co-ordinates, the Szego kernel as the function
on $S^1 \times S^1$  given by
\[
   \sigma_{(0,0)}(e^{-i\xi},e^{i\phi})=\theta_3(\phi-\xi)\,
    \theta'_1(0)[\theta_3(0)\,\theta_1(\phi-\xi)]^{-1}
      \sqrt{dw}\sqrt{d\bar{z}}
\]
where $\theta_3$ and $\theta_1$ are the classical theta functions{\footnote
{With $q=e^{-\beta/2}$,
$$\theta_1(\xi)= 2\sum_{n=0}^{\infty}(-1)^nq^{(n+1/2)^2}\sin[(n+1/2)\xi],
\quad \theta_3(\xi)= 1+2\sum_{n=1}^\infty q^{n^2}\cos (n\xi)$$}}
with $w=e^{i\phi},\,z=e^{i\xi}$ and the factor $\sqrt{dw}\sqrt{d\bar{z}}$
is a notation for a section of $L^{1/2,1/2}$. These sections
of $L^{1/2,1/2}$ are
 1/2-forms in each variable, the result of the fact that the definition
of the Szego kernel involves the so-called
prime form which is constructed from a
quotient of the theta function in the denominator by sections of
$L^{1/2,1/2}$.  Note that our notation for the theta functions is
classical and differs from that of Fay.

On the other hand in \cite{CHa}
there is a novel proof using quantum field theory
of the (well known) identity:
\[
   \Sigma_n\, e^{-\beta(n+1/2)}(1+e^{-\beta(n+1/2)})^{-1}\,e^{in(\phi-\xi)}
    = \theta_3(\phi-\xi)\,\theta'_1(0)
        [\theta_3(0)\,\theta_1(\phi-\xi)]^{-1}\, e^{i(\xi-\phi)/2}.
\]
This may be interpreted as giving the Fourier expansion of the
Szego kernel. Using this identity let us compute the effect of applying the projection
defined by integration against
the Szego kernel. To simplify the discussion we note that
$$H^{1/2,0}(\partial\bar R)\cong L^2(S^1)\oplus L^2(S^1)$$
where the first copy of $L^2(S^1)$ corresponds to the circle
$\{w:|w|=1\}$ and the second copy to the other boundary circle
of $\bar R$.
We consider an element $(f\sqrt{dz},0)\in H^{1/2,0}(\partial\bar R)$
where $f\in L^2(S^1)$
and integrate to define an extension of this section
of $L^{1/2,0}$ to a holomorphic section on the interior of $\bar R$:
$$\int \sigma_{(0,0)}(e^{-i\xi},w)f(\xi)\sqrt{dz}\equiv g(w).$$
Now we restrict this section
 $g(w)$ to $\{w:|w|w=1\}$ and to the other boundary circle
respectively to give a pair of sections
$(g_1\sqrt{dw}, g_2\sqrt{dw})$ where $g_j$ are in $L^2(S^1)$.

After some calculation
using
\[
   f(\xi)=(2\pi)^{-1}\, \Sigma_n\,\hat{f}_n\,e^{in\xi}.
\]
the end result is
\[
g_1(\phi)\sqrt{dw}=   \Sigma_n\,e^{-\beta(n+1/2)}(1+e^{-\beta(n+1/2)})^{-1}\,
     e^{in\phi}\,\hat{f}_n\sqrt{dw}
\]
\[
g_2(\phi)\sqrt{dw}= \Sigma_ne^{-\beta(n+1/2)/2}(1+e^{-\beta(n+1/2)})^{-1}\,
       e^{in\phi}\,\hat{f}_n\sqrt{dw}.
\]
The significance of this formula is that it represents the action on
$(f,0)\in L^2(S^1)\oplus L^2(S^1)$
of the projection operator $P(A(\beta))$
of Subsection~2.3.4 with $A(\beta)$ given by (19) where
$D$ in that formula is the operator of
 multiplication on the $n^{th}$ Fourier
coefficient by $n+1/2$. That is
$$P(A(\beta))\left( \begin{array}{c}
f \\
0
\end{array}\right)=
\left(\begin{array}{c}
g_1 \\
g_2
\end{array}\right).$$
As we noted in 2.3.4, $A(\beta)$ defines
the quasifree KMS state on the C*-algebra $A(L^2(S^1))$ for each
$f\in L^2(S^1)$.  This situation can be described succinctly by saying
that the effect of considering quantum field theory on a
genus one Schottky double
is to consider temperature states on the appropriate
fermion algebra.

\section{Free fields on Riemann surfaces}
\label{s8}
\subsection{Overview}

Loop groups are intimately related to conformal
field theory.
The viewpoint of Segal \cite{Seg2} gives an
axiomatic framework for conformal field theories.
Explicit examples have been constructed in \cite{CH2}
generalising the construction of the previous section.
Roughly speaking the idea is that if one has a set of oriented
circles one may `interpolate' between them using Riemann surfaces.
With each boundary circle we can associate a Hilbert
space identified with $L^2(S^1,\C^N)$
by choosing a local complex co-ordinate which
parametrises a neighbourhood of each boundary circle.
The direct sum over all boundary circles of these $L^2$ spaces
is the boundary Hilbert space.
By restricting the holomorphic sections of certain rank $N$ bundles
over the Riemann surface to the boundary one obtains
a subspace of the boundary Hilbert space.
The orthogonal projection onto this subspace may be used
to define a quasifree representation $\pi$ of the fermion algebra built
over the boundary Hilbert space.  Smooth functions from the
boundary into the structure group of the bundle on the
Riemann surface form an analogue of the loop group
(cf the torus example).  There is a representation
of this group of functions in the Hilbert space of the
quasifree fermion algebra representation $\pi$.

We will now explain this construction in more precise terms
for the case $N=1$ (which is in a sense sufficient see \cite{CH2}).
Thus we will be considering groups of maps into the group $\U(1)$
which has been the main focus throughout this article.
Consider a Riemann surface with boundary
a smooth oriented 1-manifold $S$
(which may be thought of concretely as a disjoint union of
circles).
A spin structure on a Riemann surface
 is a real line bundle $\lambda$ such that the tensor
product with itself
$\lambda\otimes\lambda$ is the cotangent bundle whose
sections are one forms on the Riemann surface.
Restricting the spin structure to $S$ we get
a real space ${\K}_{\R} = {\mathcal S}(S,\lambda)$
of smooth sections of $\lambda$.
This has a canonical quadratic form which pairs
sections $\alpha_1$ and $\alpha_2$ to give
\eq
\label{bilinear}
(\alpha_1,\alpha_2) = \int_{\partial\Sigma_1}\alpha_1\otimes\alpha_2.
\eqend
(Note that $\alpha_1\otimes\alpha_2$ is a one form and hence
can be integrated along $S$.)
Using this bilinear form we can
construct the complex Clifford $*$-algebra $C({\K}_{\R})$
over ${\K}_{\R}$ which is the associative algebra
generated by the identity $I$ and the elements of ${\K}_{\R}$
subject to the relations
$$\alpha_1\alpha_2+\alpha_2\alpha_1=(\alpha_1,\alpha_2)I.$$

It is well known that this Clifford algebra has a
unique irreducible $*$-representation with `positive energy'
(i.e. the generator of rotations has positive spectrum) for any
parametrisation of $S$.
Independence of parametrisation
 follows because diffeomorphisms
are implemented by unitaries in the
positive energy representation of the Clifford algebra
 thus giving an equivalence of
representations defined by different
parametrisations. The infinitesimal version of this
 action of the diffeomorphism group of the circle is well known: it is just
a representation of the Virasoro algebra.

Now suppose that
  $S$ is the boundary of a Riemann surface $\Sigma_1$.
 Let
 $L_1$ be a line bundle on $\Sigma_1$ whose restriction to $S$ is
$\lambda\otimes\C$.
 Then the Hilbert space ${\cal H}$ on which the
 Clifford algebra representation acts
is given by
the completion of $C({\K}_{\R})/{\cal J}$ where ${\cal J}$ is the left ideal
in the Clifford algebra generated by sections of $\lambda$
which extend over $\Sigma_1$ to holomorphic sections of $L_1$.
 We let the space of such
sections be denoted by
${\K}_1$.
(Another way to think about this representation of the Clifford
algebra is that it is the Fock representation
corresponding to the projection from ${\K}_{\R}$
to ${\K}_1$ and we will describe it more explicitly later.)

Given $\Sigma_1$ with boundary $S$ there are many ways
to `cap' the boundary circles to give a
Riemann surface without boundary.
Thinking of the example of the last subsection let us take $\Sigma_1=R$,
the annulus. Then we could glue on another annulus to form
the torus or we could think of $R$ as a cylinder
and cap the boundary with two discs to form a sphere.
More generally let us instead start with
 a Riemann surface without boundary
$\Sigma$ and a decomposition
$\Sigma=\Sigma_1\cup\Sigma_2$ into two submanifolds which intersect in their
common smooth boundary,
$\partial\Sigma_1 = S = \partial \tilde{\Sigma}_2$, and a
line bundle $L$ over $\Sigma$ such that $L\vert_{\Sigma_j} = L_j$.
The decomposition of $\Sigma$ as $\Sigma_1\cup\Sigma_2$
naturally defines a decomposition of ${\K}$ as a direct
sum of subspaces ${\K}={\K}_1\oplus{\K}_2$
which are the restrictions to the common boundary
of holomorphic sections of  $L_j$.
Each of these is isotropic with respect to the bilinear
form in Eq.\ (\ref{bilinear}) above and this data in
turn defines a Fock representation of the Clifford algebra on the exterior
algebra over ${\K}_1$.
This space is isomorphic to the irreducible $*$-representation space
${\cal H}$.

There is a geometric
way to understand this isomorphism whenever there
is an anti-holomorphic diffeomorphism  identifying $\Sigma_1$
and  $\Sigma_2$ (we say in this case that $\Sigma$ is the
Schottky double of $\Sigma_1$ and the diffeomorphism is called the
Schottky involution.) Provided the line bundle $L$ is compatible with the
Schottky involution then the
 involution may be used to define a complex conjugation on
 ${\K}_{\R}={\K}$ which gives it a natural Hilbert
 space inner product when combined with (\ref{bilinear}).
Moreover in the case of the Schottky
double it is natural to use Araki's
self-dual CAR formalism, \cite{Ar} in which
the complex conjugation on ${\K}$,
is extended to a conjugate linear involution
on  $C({\K})$ (see the next section) and this then enables us
to connect up to the fermion algebra description of
earlier sections.

In this context one has some natural generalisations of the loop
group.  Take a complex line bundle $L$ on $\Sigma$ compatible with the
Schottky involution such that the tensor product $\overline{L}\otimes
L$ is the complexification of the cotangent bundle over $S$. Then
(\ref{bilinear}) is a pairing between ${\K}_1$ and ${\K}_2 =
\overline{{\K}_1}$ and each is an isotropic subspace of ${\K} =
{\K}_1\oplus{\K}_2$.  Now introduce the group $\G$ of real analytic
$\U(1)$ valued maps on the boundary $\partial\Sigma_1$ (that is they
are the restriction to $S$ of a $\C^*$ valued function analytic in a
neighbourhood of $S$ in $\Sigma$).  This group acts on ${\K}$ by
multiplication and so defines a group of automorphisms of the Clifford
algebra $C({\K})$.  We will see in the next section that these
automorphisms are in fact {\em implementible} so that a central
extension of $\G$ has a representation $\Gamma$ on ${\cal H}$ as in
earlier sections.  (Note: at the Lie algebra level one obtains a
representation of a Heisenberg algebra thus generalising \cite{JKL}).
The main interest in conformal field theory is in the properties of
the representation of $\G$ (or more generally in the groups of smooth
compact Lie group valued functions on $\partial\Sigma_1$).  When
$\Sigma$ is a Schottky double (using standard tools of representation
theory together with results of Segal \cite{Seg1} and Carey,
Ruijsenaars and Palmer \cite{CR,CP}), this representation is cyclic
(in fact irreducible), with cyclic vector $\Omega$ say, and one may
explicitly compute the `matrix elements'
$$\ip{\Omega}{\Gamma(g)\Omega},\ \ \ \ g\in \G.$$
The resulting formulae imply those involving the tau-functions of
\cite{KNTY} at
least in certain cases.

In the geometric setting Schottky doubles are rather special. To
consider more general cases one has to work in the complex Clifford
algebra formalism not with the fermion algebra as there is no natural
involution. This departs somewhat from the main theme of this review,
however, in the next subsection we will give a brief outline of how
this theory develops. It has an interesting application to the
Landau-Lifshitz equation to which we return at the end of the paper.
{}From a more convential conformal field theory viewpoint one may
describe what we do in the next subsection as providing a geometric
interpretation of \cite{KNTY}.  We do not however attempt to derive
explicit formulae for correlation functions.  Further examples may be
found in \cite{CHM} and \cite{CHMS} while
much of the original physics literature can be traced from the
work in \cite{ABNMV,AMV,ANMV,C,DJKM,E,N,R1,R2}.

The point of particular interest in the context of this review is the
existence of a generalisation of Segal's vertex operators \cite{Seg1}
(for genus zero) to surfaces of arbitrary genus.

\subsection{Fermions}

We now make the discussion of the overview much more explicit
but at the expense of having
to assume considerable familiarity
with the theory of Riemann surfaces. A standard reference is
 for example \cite{GH}.
 Our first task is to show how splitting the Riemann
surface into two submanifolds leads to a polarisation of
our underlying  space $\mathcal K$. Thus as before
$L$ is a line bundle over a Riemann surface $\Sigma$.
As a preliminary to considering a decomposition of  $\Sigma$ into
two submanifolds we suppose that the
surface has an open covering by two sets $U_1$ and $U_2$.
Writing ${\mathcal S}(\Sigma,\O(L))$ for the global sections
of the sheaf $\O(L)$ of germs of holomorphic sections of $L$
 and $H^1(\Sigma,\O(L))$ for the first cohomology group with
coefficients in the sheaf, the Mayer-Vietoris sequence can be written as
$$0 \to {\mathcal S}(\Sigma,\O(L))\to {\mathcal S}(U_1,\O(L)) \oplus
 {\mathcal S}(U_2,\O(L))
\to {\mathcal S}(U_1\cap U_2,\O(L)) \to H^1(\Sigma,\O(L)) \to 0.$$

In the case of fermions we choose $L$ to be an even spin structure
(a square root of the cotangent bundle) for which
${\mathcal S}(\Sigma,\O(L))$ vanishes (as happens generically, \cite{Fa}).
By Serre duality $H^1(\Sigma,\O(L))$ then
also vanishes and the sequence reduces to
$$0 \to {\mathcal S}(U_1,\O(L)) \oplus {\mathcal S}(U_2,\O(L)) \to
{\mathcal S}(U_1\cap U_2,\O(L)) \to 0,$$
from which we deduce that there is a decomposition
$${\mathcal S}(U_1\cap U_2,\O(L)) = {\mathcal S}(U_1,\O(L)) \oplus
 {\mathcal S}(U_2,\O(L)).$$

Now we return to the situation
where $\Sigma_1$ and $\Sigma_2$ are closed submanifolds of
$\Sigma$ which intersect in their common smooth boundary
$$\Sigma_1\cap \Sigma_2  = \partial\Sigma_1 = \partial\Sigma_2.$$
For $j = 1$ and 2 we choose a sequence of neighbourhoods  $U_j$ which shrink
down to $\Sigma_j$, so that ${\mathcal S}(U_1\cap U_2,\O(L))$ increases to
$\K =  {\mathcal S}(\Sigma_1\cap \Sigma_2,\O(L)) =
{\mathcal S}(\partial\Sigma_1,\O(L))$.
The spaces ${\mathcal S}(U_j,\O(L))$ then
increase to give spaces $\K_j$ such that
$$\K = \K_1\oplus\K_2$$
establishing the existence of the splitting as required.

Since $L$ is a spin bundle the tensor product of sections
$\alpha_j\in {\mathcal S}(U_j,\O(L))$ gives a
 section of the cotangent bundle $K$.
Choosing an orientation of $\partial\Sigma_1$ we may integrate
$\alpha_1\otimes\alpha_2$ round the boundary to get the natural
symmetric non-degenerate bilinear form on
$\K$ given by (\ref{bilinear}).
If both sections $\alpha_1$ and $\alpha_2$ have holomorphic extensions to
$U_1$ (or $U_2$) then their product also extends and by Cauchy's theorem the
integral defining $(\alpha_1,\alpha_2)$ vanishes. From this we deduce that
${\mathcal S}(U_j,\O(L))$ and its limit $\K_j$ are isotropic, for $j = 1$ or 2.
It is easy to see that (\ref{bilinear})
 is a non-degenerate bilinear form on $\K$ and
therefore defines a pairing of the subspaces $\K_1$ and $\K_2$.

Any decomposition of an inner product space into isotropic subspaces
$$\K = \K_1\oplus\K_2,$$
gives rise to a natural representation $\Psi_{21}$ of the Clifford algebra of
$\K$ on the exterior algebra $\wedge\K_1$.
Elements $\alpha$ of $\K_1$ act by exterior multiplication,
$$\Psi_{21}(\alpha):\alpha_1\wedge\alpha_2\wedge\ldots\wedge\alpha_r
\mapsto \alpha\wedge\alpha_1\wedge\alpha_2\wedge\ldots\wedge\alpha_r ,$$
whilst elements of $\K_2$ act by inner multiplication,
$$
\Psi_{21}(\alpha):\alpha_1\wedge \alpha_2\wedge\ldots\wedge \alpha_r
\mapsto  \sum_{k=1}^r (-1)^{k-1}(\alpha,\alpha_k)\wedge \alpha_1\wedge \alpha_2\ldots
\wedge \alpha_{k-1}\wedge \alpha_{k+1}\ldots \wedge \alpha_r.
$$
(The pairing of the isotropic subspaces $\K_1$ and $\K_2$ extends to their
exterior algebras and the inner multiplication action of $\K_2$ is just the
transpose of exterior multiplication on $\wedge\K_2$.)
These conditions determine $\Psi_{21}$ and ensure the usual relations
$$
\Psi_{21}(\beta) \Psi_{21}(\alpha) + \Psi_{21}(\alpha)\Psi_{21}(\beta) = (\beta,\alpha),
$$
for all $\beta$ and $\alpha$ in ${\cal K}_1\oplus{\cal K}_2$.

For  $j = 1$ or 2, there is a cyclic vector
$\Omega_j = 1\oplus 0\oplus 0\ldots \in \wedge {\cal K}_j$, called the vacuum
vector.
With respect to the pairing of $\wedge\K_1$ and $\wedge\K_2$, $\Psi_{21}$ and
$\Psi_{12}$ are dual representations of the Clifford algebra.

We summarise the discussion above.

\vspace*{0.45cm}

{\bf \noindent Proposition I:} \label{thm8.1} {\it Associated to every decomposition
of a Riemann surface $\Sigma$ as the union of submanifolds $\Sigma_1$
and $\Sigma_2$  with common boundary and a generic even spin structure $L$ over $\Sigma$, we have the following data\\
(i) A non-degenerate bilinear form (\ref{bilinear})  on
 the real analytic sections
$\cal K$ of $L$ restricted to $\Sigma_1\cap \Sigma_2$.\\
(ii) A Fock representation of the Clifford algebra over $\cal K$
defined by (\ref{bilinear})
 on the exterior algebra over the space of sections of $L$
restricted to either $\Sigma_1$ or $\Sigma_2$. These representations
are dual to each other. }

We now introduce the (not necessarily orthogonal) projection $P_{kj}$ onto
$\K_j$ along $\K_k$,
Since $\K_1$ and $\K_2$ are
isotropic we have
$$(\beta,P_{21}\alpha) = (P_{12}\beta,P_{21}\alpha) = (P_{12}\beta,\alpha),$$
so that $P_{21}$ and $P_{12}$ are transpose maps with respect to the bilinear
form.
It follows from the definition of $\Psi_{21}$ that
$$(\Omega_2,\Psi_{21}(\beta)\Psi_{21}(\alpha)\Omega_1) =
(\beta,P_{21}\alpha).$$

Given another decomposition $\K = \K_3 \oplus\K_2$, there is
a natural map $T_{13}^2$ from $\wedge\K_3$ to $\wedge\K_1$ which
maps $\Omega_3$ to $\Omega_1$ and intertwines the Clifford algebra
representations $\Psi_{23}$ and $\Psi_{21}$, which is defined by
$$
T_{13}^2\Psi_{23}(\alpha)\Omega_3 = \Psi_{21}(\alpha)\Omega_1.
$$
This is well-defined since $\Psi_{23}(\alpha)\Omega_3$ vanishes if and only if
$\alpha$ is in the ideal generated by ${\cal K}_2$ and then
$\Psi_{21}(\alpha)\Omega_1$ vanishes too.

The normal arena for quantum field theory is a Hilbert space, which, by the
Riesz representation theorem, means that there is an antilinear identification
of the space and its dual.
Such an antilinear map arises naturally from the geometry if one
takes $\Sigma$ to be a Schottky double (cf \cite{JKL, CH1})
with
its  natural antiholomorphic involution taking
$z\in\Sigma_1$ to the corresponding point $\tilde{z}$ in $\Sigma_2$
(thus fixing each point of the boundary). Thus, as a real manifold,
$\Sigma_2$ is
an oppositely oriented copy of $\Sigma_1$. For more on
 Schottky doubles, see \cite{Fa, H}.

\vspace*{0.45cm}

{\bf \noindent Proposition II:} \label{thm8.2} {\it Let $\Sigma$ be a Schottky double.\\
(i)  The Schottky involution induces maps of forms and $1\over2$-
forms, written for brevity as $\alpha(z) \mapsto \alpha(\tilde{z})$.
The image is an antiholomorphic $1\over 2$-form, so that its complex conjugate
is holomorphic.\\
(ii) Defining
$\tilde{\alpha}(z) = \overline{\alpha(\tilde{z})},$
we obtain an antilinear map, $\sim$ with
$$(\tilde{\alpha},\tilde{\beta}) =
\int_{\partial\Sigma_1}\overline{\alpha(z)\beta(z)} =
\overline{(\alpha,\beta)},$$
(i.e. $\sim$ is antiorthogonal).\\
(iii) The map in (ii) satisfies
$(\tilde{\alpha},\alpha) = \int_{\partial\Sigma_1} |\alpha|^2,$
and hence
$\ip{\alpha}{\beta} = (\tilde{\alpha},\beta)$
defines an inner product on $\cal K$.\\
(iv) There is a natural isomorphism of the Clifford algebra over
$\cal K$ with the fermion algebra
 over ${\cal K}_1$
(regarded as a pre-Hilbert space in the inner product in (iii)) given
by $$a(\alpha)^\ast=\Psi_{21}(\alpha)\quad \alpha\in {\cal K}_1$$
where we use the notation of 2.3.1 for the fermions. }

We use \cite{Fa} for the notation on theta functions employed in the
next result which generalises the discussion for the torus.

\vspace*{0.45cm}

{\bf \noindent Lemma.} \label{thm8.3} {\it[CHM] The projection $P_1$
onto the first component in ${\cal K}={\cal K}_1\oplus{\cal K}_2$
 is given by an integral operator.
Its kernel is the Szeg\"o kernel, $\Lambda$, which can be written explicitly
in terms of the theta function $\theta[e]$ associated to the same even
half-period $e$ which specifies the choice of spin bundle $L$, and the
Schottky-Klein prime form $E$, which is a $-{1\over 2}$-form in each of its
arguments:
$$
\Lambda(x,y) = \frac {\theta[e](y - x)}{2\pi i\theta[e](0)E(y,x)}.
$$ }

(Actually this formula makes it
 clear that $\Lambda$ can be defined for any surface,
$\Sigma$ whether or not it is a Schottky double. For the proof we
refer to \cite{CH2})

\subsection{Equivalence of representations}

There is a special case of the preceding situation for which
more detailed information is available.
Henceforth we assume, following Segal \cite{Seg2},
that the boundary of $\Sigma_1$
consists of parametrised circles (we make this assumption precise
in our next result).

\vspace*{0.45cm}

{\bf \noindent Lemma.} \label{thm8.4}
{\it Assume there are coordinate charts containing each boundary circle such that  in terms of a local coordinate $z$,
 $|z| = 1$ is the boundary circle.
Then the Hilbert space representations of the CAR defined by different
Riemann surfaces $\Sigma_1$ and spin bundles $L$ which have the same boundary
$\partial\Sigma_1$ and restriction $L|_{\partial\Sigma_1}$ which we constructed in
in Proposition I in Subsection~\ref{thm8.1} are all equivalent.}

\vspace*{0.45cm}
This is proved in \cite{CH2} using
the method of \cite{PS}, Section~8.11. It is enough for our discussion
here to
understand the polarisation.
The spin bundle
can be trivialised in such a way that its sections can be
identified either with functions on the circle or with functions multiplied by
$z^{1/2}$.
Thus square-integrable sections are either identified with $L^2(S^1)$ or with
$z^{1/2}L^2(S^1)$.
Just as there is a standard polarisation of $L^2(S^1)$ into the two Hardy
spaces $H_+$ and $H_-$, so $z^{1/2}L^2(S^1)$ can be polarised into
$z^{1/2}H_+$ and $z^{1/2}H_-$.
Then the representation defined on ${\cal F}_1$ using the
decomposition into holomorphic sections on $\Sigma_1$ and its reflection
$\Sigma_4 = \phi(\Sigma_1)$ is equivalent to that defined by using the
appropriate Hardy space decomposition of the sections of $L|_{\partial\Sigma_1}$,
and so all such representations are equivalent.

\vspace*{0.45cm}

\noindent {\bf Remark}: Having established this equivalence with the
standard representation one knows (see \cite{PS}) that the existence of the
 equivalence
does not depend on the precise choice of holomorphic local coordinate
as the group Diff$(S^1)$ acts in the Hilbert space of this standard
representation (enabling us to change parametrisation).

\vspace*{0.45cm}
In the physics literature on conformal field theory
it is not usual to assume that the
Riemann surface is a Schottky double and in fact
fermion correlation functions are written down for many examples.
A case of particular interest
is the class of representations defined by the Krichever map (see \cite{PS})
for a discussion of the latter). To understand what these correlation functions
mean we need to extend the discussion of the previous subsection.

To this end let us now compare the theory obtained by capping $\Sigma_1$ by its
Schottky dual $\Sigma_2$ with that obtained when one caps it with another space
$\Sigma_-$ to give a closed surface.
To do this we need to suppose that $\Sigma_-$ is the Schottky dual of $\Sigma_+$.
We now have three different ways of decomposing $\K$:
$$\K = \K_1\oplus\K_2 = \K_1\oplus\K_- = \K_+\oplus\K_-.$$ The first
and third of these define Fock representations $\Psi_+$ and $\Psi_1$
which, by the lemma in Subsection~\ref{thm8.4}, are intertwined by
some unitary operator $U$.  Denoting transpose with respect to our
bilinear form by $\top$ and using our earlier definitions we have
$$
(\Omega_2,\Psi_1(\alpha)U\Omega_+)
= (T_{2-}^1\Omega_-,\Psi_{21}(\alpha)U\Omega_+)$$
$$=(\Omega_-,(T_{2-}^1)^\top\Psi_{12}(\alpha)^\top U\Omega_+)\\
=(\Omega_-,\Psi_{1-}(\alpha)^\top(T_{2-}^1)^\top UT_{+1}^2\Omega_1).$$
Now, from the earlier equivalences, $(T_{2-}^1)^\top UT_{+1}^2$ intertwines
$\Psi_{-1}=\Psi_{1-}^\top$ with itself and so, by Schur's Lemma, is a multiple,
$k$,  of the identity, giving
$$(\Omega_2,\Psi_1(\alpha)U\Omega_+) = k(\Omega_-,\Psi_{-1}(\alpha)\Omega_1).$$
Applying the same argument to products in the Clifford algebra now gives
$$
\ip{\Omega_1}{\Psi_1(\alpha)\Psi_1(\beta)U\Omega_+}
= (\Omega_2,\Psi_1(\alpha)\Psi_1(\beta)U\Omega_+)\\
= k(\Omega_-,\Psi_{-1}(\alpha)\Psi_{-1}(\beta)\Omega_1)\\
=k(\alpha,P_{-1}\beta).$$
Thus correlation functions involving two different Fock cyclic vectors,
$\Omega_1$ and $\Omega_+$, can also be computed purely in terms of the
geometrical projection involving the surface obtained by capping $\Sigma_1$
with the Schottky dual of $\Sigma_+$.

This provides an interpretation of the correlation functions involved in
the Krichever construction where one uses for $\Sigma_-$ a union
of discs. In some papers these correlation functions are misleadingly
written as inner products involving the same cyclic vector (i.e.
$\Omega_1$ is identified with $\Omega_+$). Under the construction
we have given here this may only be done in the Schottky double case.
This is what distinguishes the latter from other possibilities:
it is only in the Schottky case that there is a geometrically defined inner product on $\K$ in terms of which the correlation functions are positive
definite and hence one can obtain a quantum field theory that satisfies the
Wightman axioms.

\subsection{The boson fermion correspondence}

The discussion in the previous subsection has a simple consequence.

\vspace*{0.45cm}

{\bf \noindent Corollary:}\label{8.5}
{\it The representation defined by any Riemann surface $\Sigma_1$ is
equivalent to that obtained simply by capping the $p+1$ circles which make up
$\partial\Sigma_1$ by discs, that is, it is equivalent to a tensor product of
$p+1$ standard fermion representations for a single circle.}

\vspace*{0.45cm}
In the remaining sections we shall take $\Sigma$ to be the Schottky double
formed from $\Sigma_1$ and with parametrised boundary circles.
The representation of the Clifford algebra over the space $\mathcal K$
we denote by $\Psi$ for short. (Note that although in this case $\Psi$
can be regarded as a representation of the fermion algebra we will
persist with this Clifford notation.)
The group $\G\equiv{\rm Map}(\partial\Sigma_1,{\U(1)})$ of smooth functions
from $\partial\Sigma_1$ to the complex numbers of modulus 1 acts unitarily
by pointwise multiplication on $\K = L^2(\partial\Sigma_1)$.
This group also acts as automorphisms of the Clifford algebra: for
$\xi\in\G$ and $\alpha \in \K$ we have
$$\xi: \Psi(\alpha) \mapsto \Psi(\xi\cdot\alpha). $$
We will refer to $\G$ somewhat loosely as the `bosons' even though
strictly speaking it is the Lie algebra of this group which can be
given the structure of a Heisenberg algebra and hence may be regarded as
representing bosons.

As $\G$ is the product of groups of smooth maps on each connected
component of the boundary and the representation $\Psi$ is equivalent to the
standard
representation obtained by capping the boundary by discs, this
automorphism is implemented by an irreducible projective representation
$\Gamma$ with 2-cocycle $\sigma$, that is
$$
\Psi(\xi\cdot\alpha) = \Gamma(\xi)\Psi(\alpha)\Gamma(\xi)^{-1} $$
$$
\Gamma(\xi_1)\Gamma(\xi_2) =\sigma(\xi_1,\xi_2)\Gamma(\xi_1\xi_2).$$

Choose a base point $c_j$
on the $j$-th boundary circle for $j=0,1,2,\ldots, p$.
Let $\xi\in {\G}$ and $\xi= e^{if}$. Define
$\Delta_jf$ to be the change in the value of a function $f$ after one
circuit of that boundary circle. The Lie algebra of $\G$ consists of
those $f$ with $\Delta_jf = 0$ for all $j=0,1,2\ldots, p$.

Following Segal \cite{Seg1} we note that for each $\xi\in \G$
there is a choice of
unitary $\Gamma(\xi)$ such that the cocycle has the form
$$\sigma(e^{if_1},e^{if_2}) = e^{-is(f,g)/4\pi}
=\exp\left(-{i\over 4\pi}\left(\int_{\partial\Sigma_+}f_2df_1 +
\sum_{j=1}^p f_2(c_j)\Delta_jf_1\right)\right).  $$

  Although we have not chosen to do so here this cocycle may
be derived by a purely geometric argument as in \cite{CHM}.
On the Lie algebra of $\G$ the bilinear form $s$ in the expression for
$\sigma$ is symplectic,
equipping this Lie algebra with the structure of an infinite
dimensional Heisenberg algebra.

The representation $\Gamma$ has a number of interesting properties
for a discussion of which we refer the reader to \cite{CH2}.
We restrict the exposition here to describing the boson-fermion correspondence
on $\Sigma$.

For each boundary component it follows from the corollary at the
begnning of this subsection that the fermion representation can be
recovered from the boson operators (this is the boson-fermion
correspondence of Subsection~2.6).  Explicitly we choose an annular
neighbourhood of a boundary circle and a local coordinate $z$ such
that the boundary is $|z| = 1$.  As we showed in Section~\ref{s2}
following \cite{PS, CHu, CR} the fermions can be reconstructed using
the special loops or \lq blips\rq\ $\gamma_a \in \G$ where $|a| < 1$
and
$$\gamma_a(z) = \left(\overline{a}\over a\right)^{\frac 12}
\left(\frac{z-a}{z\overline{a}-1}\right).$$
In fact, it is known that for $\alpha$ concentrated on the
annular neighbourhood of that boundary circle
$$\Psi(\alpha) = \lim_{r\to 1}\int_{|a|=r}\alpha(a)\left((1-|a|^2)^{-\frac
12}\Gamma(\gamma_a)\right)\sqrt{\frac {da}a}.$$
where the the limit means strong convergence on a dense domain (see \cite{CR}
for details).
Note that since $\alpha$ is a half-form the factor $\sqrt{da/a}$ turns this
into a form which can then be integrated.
We shall now show that these local blips can also be interpreted
on the Riemann surface in the sense that they
may be extended holomorphically to the whole of $\Sigma_1$
so that the implementers of the blips are
regularised vertex operators on the Riemann surface.

First let us consider the renormalisation factor $(1-|a|^2)^{-\frac 12}$.
In terms of the  local coordinate the prime form can be expressed as
$$E(x,y) = \frac{y-x}{\sqrt{dxdy}} + O((x-y)^3), $$
(\cite{Fa} Cor 2.5).
The local expression for the involution is $\tilde{a} = 1/\overline{a}$, so
that we have
$$E(\tilde{a},a)^{-1} = \frac{\sqrt{d(1/\overline{a})da}}{\overline{a}^{-1}-a}
= \frac{\sqrt{-d\overline{a}da}}{(1-|a|^2)},$$
whose square root is almost what we want.

Now this is not quite well defined in general, and one should rather use the
expression
$$\frac{i\theta(e-\tilde{a}+a)}{\theta(e)E(\tilde{a},a)},$$
for $e$ a half period fixed by the Schottky involution and having
prescribed behaviour
on certain cycles in $\Sigma_1$.
This period is chosen
to have the same limiting behaviour near the boundary and is, as Fay shows
(\cite{Fa} Cor 6.15 {\it et seq}), a positive section of the
bundle $|K|\otimes (2{\rm Re}(e))$ where Re$(e)$ denotes the line bundle
associated with the real part of $e$.
Considering for the moment the case when $e=0$, we see that we may take a
positive square root as a section of $|K|^{\frac 12}$.
Since the boundary is oriented the restrictions of $|K|$ and $K$ there can be
naturally identified.
The  general case of non-zero $e$ can be handled by multiplying through by an
appropriate $\exp(-\sum e_k\int \omega^0_k)$ to convert it to the previous
case.
One then ends up with a square root which is a section of
$|K|^{\frac 12}\otimes e$ restricted to an annular region containing the
boundary circle.
It converges to the
half form $(id\theta)^{\frac 12} = (dw/w)^{\frac 12}$ on the circle.
It follows that in the annular region $\alpha(a)E(\tilde{a},a)^{-\frac 12}$
is a one-form and we have:
$$\Psi(\alpha) =
\lim_{\lambda\to 1}\int_{|a|=\lambda}\alpha(a)E(\tilde{a},a)^{-\frac
12}\Gamma(\gamma_a).$$

To interpret the blip $\gamma_a$ we recall that there is a family of
distinguished meromorphic functions on $\Sigma_1$ whose values on
$\partial\Sigma_1$ have modulus 1 and with the minimal number of zeroes,
\cite{Fa}
Theorem 6.6.
Tailoring the result to our needs, for $a\in \Sigma_1$ and $s$ a suitable even
half period we take
$$\epsilon_a(z) =
\frac{\theta(z-\tilde{a}-s)}{\theta(z-a-s)}\frac{E(z,a)}{E(z,\tilde{a})}
\exp\left(\frac 12\sum_1^p \mu_j\int_a^{\tilde{a}}\omega_j\right), $$
where $1+\mu_j$ agrees  modulo 2 with the winding number of $\epsilon_a$ round
the $j$-th boundary component $\partial_j\Sigma_1$.
Since $\epsilon_a$ has modulus 1 on the boundary circles it represents an
element of $\G$. To see that near the boundary circle it behaves in the
correct fashion we record the following fact.
\vspace*{0.45cm}

{\bf \noindent Lemma:} \label{8.6}
{\it In sufficiently small annular neighbourhoods of the boundary circles
the function $\gamma_a^{-1}\epsilon_a$ is defined and  converges pointwise
as $a\to w$ on the unit
circle to the constant function 1.}

\vspace*{0.45cm} One way to interpret this lemma is that as $a\mapsto
w$ on the unit circle the meromorphic function $\epsilon_a$ on the
Riemann surface approaches a (singular) distribution.  Recalling
Segal's blip construction of vertex operators described in
Section~\ref{s2} (see \cite{PS}) this suggests how to construct
regularised `vertex operators' on the Riemann surface which give a
precise analytic meaning to the boson-fermion correspondence. With
some extra work one now proves \cite{CH2}:

\vspace*{0.45cm}

{\bf \noindent Theorem:} \label{thm8.7} {\it For $\Phi$ in a  dense domain of the Fock space
$$\Psi(\alpha)\Phi = \lim_{\lambda\to 1}\int_{|a|=
\lambda}\alpha(a)\left(E(\tilde{a},a)^{-\frac
12}\Gamma(\epsilon_a)\right)\Phi.$$ }

\subsection{The Landau-Lifshitz equation}

In the previous subsection we emphasised the situation when the
splitting of the Riemann surface is into the two halves of a
 Schottky double. When the decomposition of the Riemann surface
is not symmetrical then there are additional complexities.
These are illustrated very clearly in the application
of the general ideas of the preceding subsections to the
completely integrable non-linear
Landau-Lifshitz (LL) equation in \cite{CHMS}. The interest in this
example stems from the fact
that the spectral curve $\Sigma$ which arises from the Lax form of the
LL equations is
an elliptic curve.  This means that there is no immediate
generalisation of the methods
of solving integrable systems which are applicable when the spectral
curve is the Riemann sphere. For example,
a key ingredient in the Riemann sphere case is that a generic ${\rm
SL}_2(\C)$-valued loop on the unit circle has a Birkhoff factorisation
as a product of loops, one holomorphic inside the unit disc, the other
holomorphic outside the disc (see \cite{PS} for
a discussion of Birkhoff factorisation).
 There is, however, no such factorisation
in general for a disc in $\Sigma$.

The approach to the study of the LL equation in \cite{CHMS} was partly
 modelled on the study of the KdV equation in \cite{SW}. The first
 step is to find the appropriate group of functions on the elliptic
 curve to play the role that the loop group of SL$_2(\C)$ does for
 integrable systems such as KdV with spectral curve the Riemann
 sphere.  The group was constructed by first decomposing the elliptic
 curve (or torus) into two submanifolds. The first submanifold,
 $\Sigma_1$, is a union of four disjoint discs and the second is the
 closure of the complement of $\Sigma_1$.  If we regard the torus as
 the complex plane modulo a lattice $2\pi\Z+\tau\Z$ as in
 Section~\ref{s7} then the four discs are centred on the points
 $0,\pi,\tau/2,\tau/2+\pi$. The boundary of $\Sigma_1$ consists of
 four circles and the group of interest in this case consists of the
 real analytic maps of these four circles into SL$_2(\C)$ which are
 equivariant under the discrete symmetries of the torus generated by
 translation through the half periods $\pi,\tau/2$ of the underlying
 lattice in $\C$.  The discs are clearly permuted by this symmetry
 group. The imposition of equivariance under the discrete symmetry
 quite remarkably turns out to be exactly what is needed for the
 existence of an appropriate analogue of Birkhoff factorisation.  This
 factorisation then enables one to construct the soliton solutions of
 LL by a method that generalises that in \cite{SW}(see \cite{CHMS}).

\end{document}